\documentclass[11pt]{article}
\usepackage[latin1]{inputenc}
\usepackage{amsmath}
\usepackage{amsfonts}
\usepackage{amssymb}
\usepackage{graphicx}
\usepackage{footnote}
\usepackage{float}

\usepackage{subfigure}

\DeclareGraphicsExtensions{.bmp,.png,.pdf,.jpg,.eps}

\usepackage[colorlinks]{hyperref}
\hypersetup{
    colorlinks=true,
    linkcolor=blue,
    filecolor=magenta,  
    urlcolor=[rgb]{0.00,0.00,0.50},    
    citecolor=red,
    }

\usepackage{url}

\advance \textheight by \topskip
\usepackage{amsmath}
\usepackage[english]{babel}
\makeatletter
 \@addtoreset{equation}{section}
\makeatother

\usepackage{amsmath}
\numberwithin{equation}{section}

\advance \textheight by \topskip
\usepackage{amsmath}
\usepackage[english]{babel}

\def\be{\begin{equation}}
\def\ee{\end{equation}}

\def\R{\mathbb R}
\def\W{\mathbb W}

\def\C{\mathbb C}

\usepackage{todonotes}

\begin{document}

\title{
{\bf 
Perfectly invisible 
$\mathcal{PT}$-symmetric zero-gap systems, conformal field theoretical kinks, 
 and exotic \\
 nonlinear supersymmetry
 }}

\author{{\bf Juan Mateos Guilarte${}^a$ and Mikhail S. Plyushchay${}^b$} 
 \\
[8pt]
{\small \textit{
${}^a$Departamento de F\'{\i}sica Fundamental and IUFFyM, Universidad de Salamanca,}}\\
{\small \textit{ Salamanca E-37008, Spain}}\\
{\small \textit{ ${}^b$Departamento de F\'{\i}sica,
Universidad de Santiago de Chile, Casilla 307, Santiago 2,
Chile  }}\\
[4pt]
 \sl{\small{E-mails:  \textcolor{blue}{guilarte@usal.es}, 
\textcolor{blue}{mikhail.plyushchay@usach.cl}
}}
}
\date{}
\maketitle

\begin{abstract}
We investigate  a special class of  the $\mathcal{PT}$-symmetric
quantum models  being  
perfectly invisible zero-gap systems
with a unique bound state at  the very edge
of continuous spectrum of scattering states. 
The family includes  the $\mathcal{PT}$-regularized two particle 
Calogero systems (conformal quantum mechanics models
of  de Alfaro-Fubini-Furlan) and their rational extensions
whose potentials  satisfy equations of the KdV hierarchy
and   exhibit, particularly, a behaviour typical for
extreme waves.
We show  that the two simplest Hamiltonians 
from the Calogero subfamily 
determine the fluctuation spectra around the
$\mathcal{PT}$-regularized 
kinks arising as traveling waves in the field-theoretical 
Liouville and $SU(3)$  conformal Toda systems. 
Peculiar properties of the 
quantum systems 
are reflected 
in the associated exotic nonlinear supersymmetry  
in the unbroken or partially broken phases.
The  conventional  $\mathcal{N}=2$ supersymmetry 
is extended here to the   $\mathcal{N}=4$  nonlinear supersymmetry
that involves  two bosonic generators composed from
Lax-Novikov integrals 
of the subsystems, one of which is the central charge
of the superalgebra.
Jordan states are shown to play an essential role in the construction.
\\
\end{abstract}

\vskip.5cm\noindent

\section{Introduction}\label{SecIntro}
$\mathcal{PT}$-symmetric quantum systems  \cite{BenBoe}
have many interesting properties
and attract a lot  of attention,
for reviews see \cite{Bender,Mostafa}.
Some time ago there was revealed their 
close relationship with certain two-dimensional 
integrable quantum field theories
in the context of the ODE/IM correspondence
 \cite{DorDunTat0,DorDunTat1}. 
On the other hand, many works were devoted 
to investigation of diverse aspects of the 
$\mathcal{PT}$-symmetric regularizations, extensions and 
 deformations of Calogero systems 
 \cite{Calogero,ZnoTat,GhoGup,BasMal,FriZno,FriSmi,Fring,CorrLech}.
 Calogero systems, as is well known,  govern the dynamics of the moving 
 poles  of rational solutions to the 
 Korteweg-de Vries 
 (KdV)
  equation \cite{AirMcKMos,AdlerMoser,GorNek}.
$\mathcal{PT}$-symmetric quantum systems
 also reveal peculiar  properties 
 from the perspective of supersymmetric quantum
 mechanics \cite{ZCBR,Znojil,DorDunTat2,BagMalQue,CorPly1+,CorPlyu1}.

Solitonic (reflectionless)  and finite-gap periodic quantum systems and  their 
rational, singular on real line,  limit cases are intimately related 
to the KdV equation via the inverse scattering transform. 
The KdV equation (as well as equations of the KdV  hierarchy)  appears 
as a compatibility condition of the overdetermined system of 
two equations  imposed on one  wave function. 
One of them  has a form of the stationary 
quantum Schr\"odinger equation with a spectral parameter.
Another 
defines the   evolution of the wave function in a 
 time variable on which 
 potential  depends as on  external parameter corresponding to 
 an isospectral deformation of the  Schr\"odinger equation.
Such reformulation of  the KdV equation,
introduced by Lax, naturally explains  the appearance 
of the inverse scattering transform  under construction of  
soliton solutions. The Lax representation also plays a fundamental role
in algebro-geometric method developed for 
the construction of finite-gap (finite-zone) periodic and quasi-periodic 
solutions \cite{NMPZ}.
The 
overdertermined system of equations 
in the Lax representation 
is covariant under
Darboux transformations. 
This covariance allows to employ the Darboux  transformations
as alternative method 
for construction of both stationary and time-dependent solutions to 
the KdV equation and equations of the hierarchy \cite{MatSal}.
Darboux  transformations can also be used for construction
of time-dependent  self-similar 
rational solutions to the KdV equation \cite{BorMat,MatPositons}.
The corresponding potentials in the last case  are
of  the type of potentials of the Calogero model
being singular functions with a double pole on the real line.
Any  solution of the KdV equation 
satisfies the ordinary differential equation known as 
Novikov equation which is  
a higher stationary KdV equation  \cite{NMPZ}. 
From the viewpoint of the 
associated Schr\"odinger operator, such  higher stationary 
KdV equation can be reinterpreted in terms of
a nontrivial Lax-Novikov integral. It is this 
higher
order differential operator that 
detects all the bound states in the associated Schr\"odinger 
quantum system
with a soliton potential, or the edge-states 
of the valence and conduction bands in the spectrum
of a system with periodic finite-gap potential,
and both  bound  and edge states in the 
solutions to the KdV equation with solitons in a
stationary asymptotically periodic background \cite{APChiral}.
This operator also separates the deformed 
plane waves of opposite chirality in the reflectionless systems 
with soliton potentials, and Bloch states with opposite values 
of quasi-momentum inside the valence and conduction bands 
in finite-gap systems. The Lax-Novikov higher order
differential operator also exists  in any
quantum  system with the KdV rational  potential
of the Calogero type. However,  in rational  case 
it is a formally  commuting with Hamiltonian operator which has
a non-physical nature due to a singularity of the potential\,:
acting on physical quantum states which nullify at the potential's
pole, it   transforms them into non-physical wave functions 
singular  at the position of the pole  \cite{COPl}.

\vskip0.1cm
In the present paper we construct and  investigate a special class
of the perfectly invisible $\mathcal{PT}$-symmetric 
zero-gap quantum mechanical systems 
related to the KdV hierarchy. They represent 
 the 
$\mathcal{PT}$-regularized two-particle Calogero
systems (conformal quantum mechanics models
of  de Alfaro-Fubini-Furlan \cite{deAFF}) and their rational extensions,
 which have the unique
bound state at the very edge of the 
continuous spectrum of scattering states, and
are characterized  by transmission amplitude 
equal to one. 
\vskip0.1cm

The $\mathcal{PT}$-symmetric regularization is achieved here
via a simple imaginary shift of the coordinate,  
$x\rightarrow x+i\alpha$, $\alpha\in \R$~\footnote{
For 
$\mathcal{PT}$-symmetric deformations of the KdV and other
integrable systems in spirit of
\cite{BenBoe}, see refs. \cite{BBCF,FriKDV}
and \cite{BenFei,CavFring}.}.
As a result, the  Lax-Novikov  
operator will transform into a true integral of the
corresponding quantum system.
The real and imaginary parts of the potential
with `reconstructed' time-dependence
will provide us with interesting 
solutions to the KdV  and higher equations of the hierarchy
which exhibit, particularly,  the behaviour typical for
the extreme (rogue) waves. 
The corresponding potentials  are constructed
via Darboux transformations which use 
as the seed states a complex non-physical eigenstate 
of zero energy of a free
quantum particle and  the Jordan states related to 
it.
As a consequence, each of the obtained quantum systems 
will possess  a unique bound state of zero energy 
at the very edge of the continuous part of the 
spectrum. Since the quantum $\mathcal{PT}$-symmetric 
systems are generated from 
the free particle, and their  potentials are 
non-singular on the real line, the obtained systems will be 
reflectionless. 
In the case of usual  reflectionless systems with soliton 
potential, the transmission amplitude is a pure phase
which is a rational function of energy  of 
scattering states with zeroes and poles 
defined by energies of the bound states. 
In the the quantum 
$\mathcal{PT}$-symmetric systems we study here,
there is a unique bound state of zero energy
that is located at the very edge of the continuous part
of the spectrum.
As a consequence, the transmission amplitude 
does not depend on energy of the scattering states
and reduces   to a constant value equal to $1$. 
This means that the systems we consider 
are not just  reflectionless, but are 
perfectly invisible systems. Such quantum systems are studied,
particularly,  
in the context of quantum optics 
\cite{LonDel,CJP,KYZ}.
The described peculiarities 
of the perfectly invisible
zero-gap quantum systems lead to
unusual properties of the corresponding 
supersymmetrically extended systems.
Because of the presence of the nontrivial 
Lax-Novikov integrals of motion,
instead of the conventional  $\mathcal{N}=2$ supersymmetry, 
the Darboux-related quantum pairs 
will be described by exotic  $\mathcal{N}=4$ nonlinear  supersymmetry.
The anti-commutators of 
linear and higher order supercharges appearing here 
generate the Lax-Novikov integral of the extended system,
which plays a role of the bosonic central charge of the superalgebra.
We also will show  that the two simplest Hamiltonians 
from the Calogero subfamily 
determine the fluctuation spectra around the
$\mathcal{PT}$-regularized 
kinks arising as traveling waves in the field-theoretical 
Liouville and $SU(3)$  conformal Toda systems.  

The paper is organized as follows.
In the next Section \ref{Section2} we construct the indicated class of the 
quantum systems 
by applying the appropriate Darboux-Crum transformations to a free 
particle. We investigate their relationship with the stationary and non-stationary 
equations of the KdV hierarchy, and describe the properties 
of the higher-derivative Lax-Novikov integrals in such systems.
We also consider there some particular 
time-dependent $\mathcal{PT}$-symmetric 
 potentials  whose real and imaginary parts
have a soliton-like form with a behaviour
typical for extreme waves. 
In Section \ref{Section3} we study 
$(1+1)$-dimensional conformal field theoretical kinks 
appearing in the 
Liouville and $SU(3)$  conformal Toda systems 
and establish their relation with 
two simplest cases of  perfectly invisible
$\mathcal{PT}$-symmetric Calogero systems.
In Section \ref{Section4}  we   discuss
 exotic $\mathcal{N}=4$ nonlinear supersymmetry 
of the extended systems composed from the pairs 
of $\mathcal{PT}$-symmetric zero-gap systems
related by the first order Darboux transformations.
Section \ref{Section5} is devoted to the summary, discussion and outlook.
In Appendix we briefly discuss a 
quantum scattering problem on the half-line.

\section{Perfectly invisible $\mathcal{PT}$-symmetric zero-gap systems}\label{Section2}
\subsection{$\mathcal{PT}$-regularized Calogero  systems}

Consider a  quantum free particle on $\R^1$ described by the
Hamiltonian 
 $H_0=-\frac{d^2}{dx^2}$.
Eigenstates of $H_0$ are the plane waves $\psi_{0,\pm k}=e^{\pm
ikx}$, $k>0$, and any positive energy value $E_k=k^2>0$ is doubly
degenerate. A non-degenerate eigenstate $\psi_{0,0}(x)=1$ of zero  energy
corresponds to a limit $k\rightarrow 0$ case of the plane waves
$\psi_{0,\pm k}$. Like
$\psi_{0,\pm k}$, $\psi_{0,0}$ is a bounded function on $\R^1$.
A  linear  independent from   $\psi_{0,0}(x)$ eigenstate of
$H_0$ of zero eigenvalue is  $\widetilde{\psi_{0,0}}=x$,
 that is a non-physical state unbounded at infinity.
 In general,
a linear independent from $\psi(x)$ solution to a
stationary Schr\"odinger equation $H\psi=E\psi$
can be taken in the form
\begin{equation}\label{psitilde}
\widetilde{\psi(x)}=\psi(x)\int \frac{d\xi}{(\psi(\xi))^2}\,.
\end{equation}
Normalization  in (\ref{psitilde}) corresponds to
the Wronskian value $W(\psi,\widetilde{\psi})=1$.
The state  $\widetilde{\psi_{0,0}}=x$ also
can be obtained  from  the odd
 linear combination of the plane wave
states of energy $k^2$ in a
limit $k\rightarrow0$\,:  $\lim_{k\rightarrow
0}(\psi_{0,+k}-\psi_{0,-k})/2ik=\widetilde{\psi_{0,0}}$.

The Hamiltonian operator $H_0$ has an
integral of motion $\mathcal{P}_0=-i\frac{d}{dx}$,
$[H_0,\mathcal{P}_0]=0$. It
distinguishes the left- and right-moving
plane wave states inside the energy doublets,
$\mathcal{P}_0\psi_{0,\pm k}=\pm k \psi_{0,\pm k}$.
The state
$\psi_{0,0}=1$ constitutes the kernel of $\mathcal{P}_0$,
$\mathcal{P}_0\psi_{0,0}=0$, while
non-physical zero energy
eigenstate $\widetilde{\psi_{0,0}}$ is transformed
by  $\mathcal{P}_0$ into the eigenstate $\psi_{0,0}$,
 $i\mathcal{P}_0 x=1$. As
$(\mathcal{P}_0)^2\,\widetilde{\psi_{0,0}}=0$,
the state $\widetilde{\psi_{0,0}}$ is
a Jordan state of
$\mathcal{P}_0$ of order $2$ \cite{CarPly}.

Within a framework of Darboux transformations, one can construct
the annihilator of the non-physical zero energy state $\widetilde{\psi_{0,0}}(x)=x$\,:
$D_1=x\frac{d}{dx}x^{-1}=\frac{d}{dx}-\frac{1}{x}$, $D_1x=0$. 
The conjugate operator is $D_1^\dagger=-x^{-1}\frac{d}{dx}x=-\frac{d}{dx}-\frac{1}{x}$,
and these two operators provide a factoirzation of $H_0$\,:
\begin{equation}\label{H0factor}
    H_0=D_1^\dagger D_1=H_0\,.
\end{equation}
Their permuted product generates a conformal quantum mechanics model
\cite{deAFF}
given by a Hamiltonian for relative motion of a 2-particle Calogero
system
\begin{equation}\label{H1factor}
    H_1=D_1D_1^\dagger=-\frac{d}{dx^2}+\frac{2}{x^2}
\end{equation}
 with a value $g=2$
of the coupling constant.
Unlike $H_0$, Hamiltonian $H_1$ is
singular at $x=0$. If $H_1$ is considered on a whole
real line punctured at the origin, the potential barrier at $x=0$
is not penetrable and the states with supports on $x<0$ and $x>0$
do not mix dynamically. The singularity of operators $H_1$,
$D_1$ and $D_1^\dagger$ is associated with a node at $x=0$ of
the non-physical zero energy eigenstate $\widetilde{\psi_{0,0}}(x)=x$
of $H_0$.

Potential $u_1=\frac{2}{x^2}$ in (\ref{H1factor}) is a stationary,
singular on $\R^1$ solution~\cite{AirMcKMos} to the
KdV equation
\begin{equation}\label{KdV}
u_t-6uu_x+u_{xxx}=0\,.
\end{equation}
It can be obtained from the
one-soliton KdV solution
\begin{equation}\label{PoTel}
u_{1}(x,t)=-\, \frac{2\kappa^2}{\cosh^2\kappa(x-4\kappa^2t-\tau)}\,
\end{equation}
by fixing   the soliton centre coordinate to be pure imaginary and equal
$\tau=i\frac{\pi}{2\kappa}$, and then by taking a limit
$\kappa\rightarrow 0$. The limit $\kappa\rightarrow 0$
eliminates a length scale which controls the
size of the one-soliton solution correlated with its velocity.
As a result,
 the stationary Schr\"odinger equation
$H_1\psi=E\psi$
is invariant under the transformation
$x\rightarrow \alpha x$, $E\rightarrow \alpha^{-2}E$,
that corresponds to a scale invariance of the system (\ref{H1factor})
similarly to the case of the free particle.
A time-dependent solution
$u_1(x,t)=-\frac{1}{6}c +\frac{2}{(x-ct)^2}$
to the KdV equation   is
obtained  from the
stationary solution $u_1(x)=\frac{2}{x^2}$
by employing the invariance
of  (\ref{KdV})   under Galilean transformations,
$u(x)\rightarrow u(x,t)=-\frac{1}{6}c +u(x-ct)$.
It is a singular solution with a moving along $\R^1$ second order pole
whose velocity $c$ is correlated with its background asymptotic value
$-\frac{1}{6}c$. The 
time-dependent solution $u_1(x,t)$ also can  be obtained 
directly from the one-soliton solution (\ref{PoTel}) 
in a more tricky way 
without making use of the Galilean symmetry. 
For this, after fixing  $\tau=i\frac{\pi}{2\kappa}$,
in the resulting intermediate singular one-soliton solution
 $u=\frac{2\kappa^2}{\sinh^2\kappa X}$
to the KdV equation
we take a limit $\kappa\rightarrow 0$
by preserving quadratic in $\kappa^2$ terms
but leaving the composed argument $X=x-4\kappa^2t$
untouched.
In such a  way we obtain $u=\frac{2}{X^2}-\frac{2}{3}\kappa^2$.
Denoting then  $4\kappa^2=c$, we
arrive at  the same time-dependent singular solution
$u_1(x,t)$.

Let us return to the Hamiltonian operators  (\ref{H0factor}) and  (\ref{H1factor}).
{}From their  factorization properties it follows that the operators
$D_1$ and $D_1^\dagger$ intertwine $H_0$ and $H_1$\,: $D_1H_0=H_1D_1$,
$H_0D_1^\dagger=D_1^\dagger H_1$. As a consequence, non-singular at
$x=0^+$ physical states of $H_1$ on half-line $x>0$ are obtained from the odd
linear combination of the plane wave  eigenfunctions  of $H_0$,
$\psi_{1,k}(x)=D_1\sin kx$. They satisfy the relations
$H_1\psi_{1,k}=k^2\psi_{1,k}$ and
$\lim_{x\rightarrow 0}\psi_{1,k}(x)=0$.
 The operator
\be\label{P1(x)}
 \mathcal{P}_1=D_1 \mathcal{P}_0
D_1^\dagger= i \left(\frac{d^3}{dx^3}-\frac{3}{x^2}\cdot\frac{d}{dx}+\frac{3}{x^3}\right),
\ee
being a Darboux-dressed free particle integral
$\mathcal{P}_0$, commutes with $H_1$.
This is a Lax-Novikov operator for the two-particle
Calogero system (\ref{H1factor}).
 It is not,
however,  a true integral of motion of $H_1$  since
 acting on physical
eigenstates $\psi_{1,k}(x)$ it transforms them into non-physical
eigenstates of $H_1$ of the same energy $E=k^2>0$
which are singular at
$x=0$\,:  $\mathcal{P}_1\psi_{1,k}(x)=-ik^3D_1\cos kx=ik^3(k\sin kx+\frac{1}{x}\cos kx)$.

The singularity at $x=0$  of the intertwining operators $D_1$ and $D_1^\dagger$
and consequently of the Hamiltonian $H_1$ and operator $\mathcal{P}_1$
can be removed by a
`$\mathcal{PT}$-regularization'.
This is achieved by taking a
complex linear combination of
non-physical and physical zero energy eigenstates of $H_0$,
\begin{equation} \label{Darbgenfun}
\widetilde{\psi_{0,0}^\alpha}=x+i\alpha\equiv \xi\,.
\end{equation}
where $\alpha$ is
a nonzero real constant which, for definiteness, will 
be taken positive, $\alpha>0$.
With the help of this state
 we construct the first order
differential operators
\begin{equation}\label{D1}
    D_1=\xi\frac{d}{dx}\frac{1}{\xi}=\frac{d}{dx}-\xi^{-1}\,,\qquad
    D_1^\#= -\frac{1}{\xi}\frac{d}{dx}\xi=-\frac{d}{dx}-
    \xi^{-1}\,,
\end{equation}
whose kernels are, respectively, $\xi$ and $\xi^{-1}$.
As before  we
have $H_0=D_1^\# D_1=-\frac{d^2}{dx^2}$.
But now the partner
Hamiltonian operator $H_1=D_1 D_1^\#=-\frac{d^2}{dx^2}+2\xi^{-2}$ is
non-singular on the real line $\R^1$,
\begin{equation}\label{H1PT}
    H_1^\alpha=-\frac{d^2}{dx^2}+\frac{2}{(x+i\alpha)^2}\,,
\end{equation}
and the first order differential operators (\ref{D1})
intertwine $H_0$ and $H_1^\alpha$,
\be\label{HaH0inter}
D_1H_0=H_1^\alpha
D_1\,, \qquad
D_1^\# H_1^\alpha=H_0 D_1^\#\,.
\ee
Potential of the system (\ref{H1PT}) is obtained
from the one-soliton solution (\ref{PoTel})
to the KdV equation if in the procedure
described above
the soliton centre parameter is fixed in the form
$\tau=\frac{i}{\kappa}(\frac{\pi}{2}-\alpha)$.

Hamiltonian (\ref{H1PT}) with a complex potential
\begin{equation}\label{VPTsym}
V_1(x)=2\frac{x^2-\alpha^2}{(x^2+\alpha^2)^2}-4i\alpha
\frac{x}{(x^2+\alpha^2)^2}\,
\end{equation}
is
${\mathcal{PT}}$-symmetric, $[H_1^\alpha,{PT}]=0$.
Here ${P}$ is a space  reflection operator, ${P}x=-x{P}$,
${P}^2=1$, and a complex conjugation operator
 ${T}$ is  defined by
 ${T}z=\bar{z}\,{T}$, ${T}^2=1$,
where $z\in \C$ is an arbitrary complex number.
If we extend the definition of
time reflection operator $T$ by
a requirement $Tt=-tT$,  the time-dependent
KdV equation (\ref{KdV}) will be invariant 
under the $\mathcal{PT}$ transformation
if $u(x,t)$ is  $\mathcal{PT}$-symmetric\,:
$[u(x,t),PT]=0$.
Then the change $x\rightarrow X=x-ct$
accompanied by a shift for $-\frac{1}{6}c$ transforms 
potential (\ref{VPTsym}) being stationary solution of the KdV equation
into $\mathcal{PT}$-symmetric time-dependent 
travelling wave solution $u(x,t)=V_1(X)-\frac{1}{6}c$ of the 
same equation.

The bounded
on real line $\R$  eigenstates 
  $\psi_{1,\pm k}^\alpha(x)=D_1e^{\pm ik\xi}$, $k>0$,
$H_1^\alpha\psi_{1,\pm k}^\alpha=k^2\psi_{1,\pm k}^\alpha$,
 can be considered as physical states
of the ${\mathcal{PT}}$-symmetric system
(\ref{H1PT}).
 They are the Darboux-transformed plane wave eigenstates of
$H_0$ that are obtained by applying to  them  the intertwining
operator $D_1$ defined in (\ref{D1}).
Unlike $H_0$, the system (\ref{H1PT})
has  a gapless bound state of zero energy  that lies at the very  edge
of the doubly degenerate continuous part of the spectrum. This
 square-integrable on $\R^1$  state  can be
obtained from the bounded but not square-integrable singlet
zero energy eigenstate $\psi_{0,0}=1$  of $H_0$ by applying to
the latter
the intertwining operator
$D_1$\,: $-D_1\psi_{0,0}=\xi^{-1}\equiv \psi_{1,0}^\alpha(x)$,
$H_1^\alpha\psi_{1,0}^\alpha=0$. The ground state $\xi^{-1}$
corresponds to a limit $k\rightarrow 0$ of  the physical states
$\psi_{1,\pm
k}^\alpha$
 from the
continuous part of the spectrum, $\lim_{k\rightarrow 0}(-\psi_{1,\pm
k}^\alpha)=\xi^{-1}$.
It also can be obtained from the
bound eigenstate 
 ${\kappa}/{\cosh\kappa (x+\tau)}$ of eigenavalue $-\kappa^2$
of the reflectionless P\"oschl-Teller system
given by potential (\ref{PoTel}) at  $t=0$ after
fixing  $\tau=\frac{i}{\kappa}(\frac{\pi}{2}-\alpha)$
and taking  a limit $\kappa\rightarrow 0$.
In this limit the energy gap separating
the bound state from the continuous part of the spectrum
disappears, and the bound eigenstate
of the P\"oschl-Teller system
with one-soliton potential (\ref{PoTel}) transforms into
the bound state of the system $H_1^\alpha$.

The peculiarity of the reflectionless $\mathcal{PT}$-symmetric system
(\ref{H1PT}) in comparison with Hermitian reflectionless systems is
that the square-integrable  on $\R^1$ singlet state here is not separated by a gap from the
continuous doubly degenerate part of the spectrum with $E=k^2>0$. In
this aspect it is similar to the periodic $\mathcal{PT}$-symmetric
finite-gap systems considered in \cite{CorPlyu1}.
It is also worth to emphasize here that the translational non-invariance 
of the $\mathcal{PT}$-symmetric system (\ref{H1PT}) generated 
from the original translation-invariant   free particle system $H_0$ is
rooted in translational non-invariance of the zero energy eigenstate
(\ref{Darbgenfun}) of $H_0$  underlying the  Darboux transformation.

Unlike (\ref{H1factor}), the
$\mathcal{PT}$-symmetric system (\ref{H1PT}) possesses a true
integral of motion being a Darboux-dressed integral $\mathcal{P}_0$ of the
free particle, $\mathcal{P}_1^\alpha=D_1 \mathcal{P}_0 D_1^\#$,
$[\mathcal{P}_1^\alpha,H_1^\alpha]=0$. This integral distinguishes
the doublet states in the continuous part of the spectrum,
$\mathcal{P}_1^\alpha\psi_{1,\pm k}^\alpha=\pm k^3\psi_{1,\pm
k}^\alpha$, and annihilates the ground state $\psi_{1,0}^\alpha=\xi^{-1}$,
$\mathcal{P}_1^\alpha\psi_{1,0}^\alpha=0$. The complete kernel of
the third order differential operator $\mathcal{P}_1^\alpha$
is
\begin{equation}\label{kerP1}
    {\rm ker}\, \mathcal{P}_1^\alpha
    =\text{span}\, \{\xi^{-1},\xi,\xi^3\}\,.
\end{equation}
It includes  non-physical  (undbounded)
states $\xi$  and $\xi^3$.  These two states
 are not eigenstates of
$H_1^\alpha$, but they belong to the  kernels of
the operators  $(H_1^\alpha)^2$
and $(H_1^\alpha)^3$, respectively:
\begin{equation}\label{H1xi}
    H_1^\alpha\xi^3=-4\xi\,,\quad
    (H_1^\alpha)\xi=2\xi^{-1}\,,\quad
   (H_1^\alpha)^2\xi=0\,,\quad
    (H_1^\alpha)^3\xi^3=0\,.
\end{equation}
Thus the states $\xi$  and $\xi^3$ are 
the Jordan states of
$H_1^\alpha$ of orders $2$ and $3$.
In correspondence with (\ref{kerP1}) and
(\ref{H1xi}), the operator $\mathcal{P}_1^\alpha$ satisfies a
supersymmetry-like relation 
\cite{CorPlyu2}
\begin{equation}\label{P1spectral}
    (\mathcal{P}_1^\alpha)^2=(H_1^\alpha)^3\,
\end{equation}
concordant  with the
Burchnall-Chaundy  theorem \cite{BurCha,Krich}.
It is a Lax-Novikov integral for
the finite-gap 
(zero-gap) $\mathcal{PT}$-symmetric  system $H_1^\alpha$.

A partner of the zero
energy ground state $\xi^{-1}$ of the system $H_1^\alpha$
given by relation (\ref{psitilde})
 is an unbounded state $\xi^2$,
$H_1^\alpha\xi^2=0$.
It does not belong to the kernel
(\ref{kerP1}) like  the
non-physical zero energy eigenstate $\widetilde{\psi_{0,0}^\alpha}=\xi$
of $H_0$
does not belong to the kernel of the
 integral $\mathcal{P}_0$.
The state $\xi^2$ is obtainable  from the appropriate
linear combination of the doublet states from the continuous part of
the spectrum of $H_1^\alpha$ in a limit $k\rightarrow 0$\,:
\begin{equation}\label{xi2lim}
    \lim_{k\rightarrow
    0}\frac{-3}{2ik^3}(\psi_{1,k}^\alpha-
    \psi_{1,-k}^\alpha)=\xi^2\,.
\end{equation}
The operator  $\mathcal{P}_1^\alpha$
transforms this non-physical zero energy state into a physical ground state,
$\mathcal{P}_1^\alpha\xi^2\propto \xi^{-1}$, that can be compared with  a
similar picture  in the case of
the free particle system.

With respect to the  Darboux transformation
generator ${D}_1$,
the pre-images of the states from the kernel
of $\mathcal{P}_1$
are $1$, $\xi^2$ and $\xi^4$\,:  ${D}_1 1=-\xi^{-1}$,
${D}_1 \xi^2=\xi$,
${D}_1 \xi^4=3\xi^3$.
Unlike the zero energy eigenstate $\psi_{0,0}=1$ of $H_0$,
other two states satisfy the relations
 $H_0\xi^4=-12\xi^2$ and $H_0\xi^2=-2$,
i.e. $\xi^2$ and $\xi^4$ are the
Jordan states of $H_0$ of orders
$2$ and $3$, respectively.
The nature of these three states related to the system  $H_0$
is similar to that of their Darboux-counterparts  (\ref{kerP1})
in the system $H_1^\alpha$, cf.   Eq.  (\ref{H1xi}).
\vskip0.2cm

The family of the systems
\begin{equation}\label{HnPT}
    H_n^\alpha=-\frac{d^2}{dx^2}+\frac{n(n+1)}
    {(x+i\alpha)^2}\,
\end{equation}
characterized by  the integer parameter $n\geq 0$ is
a hierarchy of the $\mathcal{PT}$-symmetric reflectionless
systems that  includes the free particle $H_0$ and
the system (\ref{H1PT}) as particular cases corresponding to
the values $n=0$ and $n=1$.
For neighbour members of the hierarchy, there are
the  factorization,
\begin{equation}\label{HnDDfactor}
    H_{n-1}^\alpha=D_n^\#D_n\,,\qquad
    H_{n}^\alpha=D_n D_n^\#\,,
\end{equation}
and the intertwining,
\begin{equation}\label{HnHn-1inter}
    D_nH_{n-1}^\alpha=H_{n}^\alpha D_n\,,\qquad
    D_n^\#H_{n}^\alpha=H_{n-1}^\alpha D_n^\#\,,
\end{equation}
 relations.
Here the first order operators
\begin{equation}\label{Dn}
    D_n=\xi^n\frac{d}{d\xi}\frac{1}{\xi^n}=\frac{d}{d\xi}-n\xi^{-1}\,,\qquad
    D_n^\#=-\frac{1}{\xi^n}\frac{d}{d\xi}\xi^n=-\frac{d}{d\xi}-
    n\xi^{-1}\,,\quad
    n=1,\ldots\,,
\end{equation}
are generated by  zero energy non-physical eigenstates $\xi^n$ of
$H_{n-1}^\alpha$, $H_{n-1}^\alpha\xi^{n}=0$.
The bounded but not square-integrable
physical eigenstates of $H_n^\alpha$
from the continuous part of the spectrum are obtained by applying
$D_n$ to the corresponding eigenstates of $H_{n-1}^\alpha$ of the same nature,
$\psi_{n,\pm k}^\alpha=
D_n\psi_{n-1,\pm k}^\alpha$, $k>0$. These states
can be  constructed
iteratively from the plane wave eigenstates of the free particle model,
\begin{equation}\label{psink}
    \psi_{n,\pm k}^\alpha(x)=\hat{\mathcal{D}}_n e^{\pm ik\xi}\,.
\end{equation}
Here we define differential operators of order $n$\,:
\begin{equation}\label{calDn}
    \hat{\mathcal{D}}_n=D_nD_{n-1}\ldots D_1\,,\qquad
    \hat{\mathcal{D}}_n^\#=D_1^\#\ldots D_{n-1}^\#D_n^\#\,.
\end{equation}
 The square-integrable zero energy ground
state $\psi_{n,0}^\alpha=\xi^{-n}$ of $H_{n}^\alpha$ is obtained from a singlet
zero energy ground state $\psi_{0,0}=1$ of $H_0$\,:
$\psi_{n,0}^\alpha\propto \hat{\mathcal{D}}_n1$.

The operators (\ref{calDn}) intertwine $H_{n}^\alpha$ directly with  $H_0$,
\begin{equation}\label{DnH0}
\hat{\mathcal{D}}_nH_0=H_n^\alpha
\hat{\mathcal{D}}_n,\qquad
\hat{\mathcal{D}}_n^\# H_n^\alpha=
H_0\hat{\mathcal{D}}_n^\# \,,
\end{equation}
and correspond to the Darboux-Crum
transformation of order $n$ of the free particle
based on non-physical eigenstate $\xi$ of $H_0$
and its certain $n-1$ Jordan states, see below.
They  allow us to construct a nontrivial
integral  of the system $H_{n}^\alpha$ in the form of the dressed
momentum operator of the free particle,
\begin{equation}\label{Pn}
    \mathcal{P}_n^\alpha=-i\hat{\mathcal{D}}_n\frac{d}{dx}
    \hat{\mathcal{D}}_n^\#\,,\qquad
    [\mathcal{P}_n^\alpha,H_n^\alpha]=0\,.
\end{equation}
Being differential operator of order $2n+1$, it distinguishes the
left- and  right-moving eigenstates
(\ref{psink})  of $H_n^\alpha$,
\begin{equation}\label{Pnpsikal}
\mathcal{P}_n^\alpha\psi_{n,\pm k}^\alpha=\pm k^{2n+1}\psi_{n,\pm
k}^\alpha\,,
\end{equation}
and satisfies the Burchnall-Chaundy, supersymmetry-type
relation
\begin{equation}\label{BCPH}
    (\mathcal{P}_n^\alpha)^2=(H_n^\alpha)^{2n+1}\,.
\end{equation}
Its kernel of dimension $2n+1$ is
\begin{equation}\label{kerPn}
    \text{ker}\,\mathcal{P}_n^\alpha =\text{span}\,
    \{\xi^{-n},\xi^{-n+2},\ldots,
    \xi^{3n-2},\xi^{3n}\}\,.
\end{equation}
Except the  zero energy ground state $\xi^{-n}$, all other
states from the kernel  are  not eigenstates of
$H_n^\alpha$ but  satisfy the relations
\begin{equation}\label{HnkerPn}
    H_n^\alpha\xi^{-n+2l}=2l(2n-2l+1)\xi^{-n+2l-2}\,,\qquad
    (H_n^\alpha)^{l+1}\xi^{-n+2l}=0\,,\qquad
    l=0,\ldots, 2n\,.
\end{equation}
Each of the states  $\xi^{-n+2l}$ with
$l=1,\ldots, 2n$
is  the Jordan state of
$H_n^\alpha$ of the corresponding order
$l+1$.

The partner (\ref{psitilde}) of zero energy ground state
$\psi_{n,0}^\alpha=\xi^{-n}$
of $H_n^\alpha$ is a non-physical state $\xi^{n+1}$.
It does not belong to the kernel
(\ref{kerPn}),  but the action of $\mathcal{P}_n^\alpha$
transforms it into the physical ground state $\xi^{-n}$. It can be
obtained from a linear combination
of the physical eigenstates from  the continuous part of the spectrum of
$H_n^\alpha$\,:
\begin{equation}\label{E0nonphys}
    \lim_{k\rightarrow 0}\frac{1}{2ik^{2n+1}}(\psi_{n,k}^\alpha
    -\psi_{n,-k}^\alpha)=C_n\xi^{n+1}\,,
\end{equation}
where $C_n$ is a constant coefficient.
It also can be obtained
from the plane wave solutions  of the free particle
by  employing relation (\ref{psink}).

In accordance with relations
(\ref{DnH0}) and  (\ref{calDn}),
 the potential $V_n(x)=\frac{n(n+1)}{\xi^2}$
can be presented in the form
\begin{equation}\label{VnWn}
V_n(x)=-2\frac{d^2}{dx^2}\big(\ln \W_n(\xi)\big)\,,
\end{equation}
where $\W_n(\xi):=W(\xi,\xi^3,\ldots,\xi^{2n+1})$
is the Wronskian of the states $\xi,\xi^3,\ldots,\xi^{2n+1}$.
Here the states $\xi^{2r+1}$ with $r=1,\ldots,n-1$
 are the Jordan states of $H_0$ of zero energy of the corresponding
 order\,:
$H_0 \xi^{2r+1} = -2r(2r+1)\xi^{2r-1}$, $(H_0)^{r+1} \xi^{2r+1}=0$.
Such a transformation based on Jordan
states corresponds to a confluent case
of some  Darboux-Crum transformation \cite{Sch-Hal,CJP,ConAst}.
 For the eigenstates  $\psi_{n,\pm k}^\alpha$ of $H_n^\alpha$
from (\ref{psink}) we have (up to a multiplicative constant factor)
an alternative representation\,:
\begin{equation}\label{psinWn}
\psi_{n,\pm k}^\alpha=\frac{W(\xi,\xi^3,\ldots,\xi^{2n+1},e^{\pm ik\xi})}{\W_n(\xi)}\,.
\end{equation}
The case $k=0$ is included in (\ref{psinWn}), and corresponds to generation
of the unique square-integrable zero energy
eingenstate $\psi_{n,0}^\alpha=\xi^{-n}$ of $H_{n}^\alpha$
from the eigenstate $\psi_{0,0}=1$ of $H_0$.
Note that the arguments $\xi^{2j+1}$ 
in the Wronskians also can  be presented in the form  $\xi^{2j+1}=
(-i)^{2j+1}\frac{\partial^{2j+1}}{\partial k^{2j+1}}e^{ik\xi}\vert_{k=0}$
in terms of the plane wave eigenstates of the  free particle. This corresponds
to the generalized Wronskian formula for solutions of the KdV equation
considered in ref. \cite{MatPositons}  that, in turn,  can be obtained
via confluent Darboux-Crum transformations \cite{CJP}.
\vskip0.1cm

The nature of the family of  the $\mathcal{PT}$-symmetric systems (\ref{HnPT})
is rather peculiar.
Like any finite-gap system,
each of the systems (\ref{HnPT})  possesses the corresponding Lax-Novikov integral
of motion (\ref{Pn})  that is a  differential operator of order $2n+1$.
Each of these systems  is  reflectionless\,: 
the plane wave states $e^{ikx}$ and $e^{-ikx}$ of the free particle $H_0$
are transformed into the deformed plane wave eigenstates
(\ref{psink}) of $H^\alpha_n$ which propagate to the left or to the right,
and are  distinguished by the Lax-Novikov integral,
see Eq. (\ref{Pnpsikal}).
The two indicated properties are also characteristic
for any system with a multi-soliton potential as for
their simplest one-soliton
representative (\ref{PoTel}).
Unlike the indicated  conventional reflectionless systems,
the  $\mathcal{PT}$-symmetric systems (\ref{HnPT})
are {\it perfectly invisible}. Namely,
 as follows from (\ref{calDn}) and
 (\ref{Dn}), the  eigenstates (\ref{psink}) corresponding to the
 upper sign case in the asymptotic region  $x\rightarrow -\infty$ have the
 form of the plane waves $Ce^{ikx}$.
 In the asymptotic region $x\rightarrow +\infty$ these states have
 exactly the same form of the plane waves $e^{ikx}$ multiplied by the same
constant  factor $C$.  A similar picture is valid for solutions that correspond
 to the lower sign in (\ref{psink}) and have the asymptotic form
$Ce^{-ikx}$ in both regions $x\rightarrow +\infty$ and $x\rightarrow -\infty$.
 The phase shift produced by a nontrivial
 potential in (\ref{HnPT}) is  therefore equal to zero ($\text{mod} \,2\pi$).
 The transmission amplitude  is equal to one,
 and any of these systems is indeed perfectly invisible.
\vskip0.1cm

The potential $V_n(x)=\frac{n(n+1)}{\xi^2}$
of $H_n^\alpha$ is a stationary $\mathcal{PT}$-symmetric
solution of the corresponding
higher order equation
of the   KdV hierarchy.
For instance, the potential $V_2(x)=\frac{6}{\xi^2}$
is a stationary solution to  the equation
\begin{equation}\label{KdV2}
u_t+30u^2 u_x -20 u_x u_{xx}-10uu_{xxx}+u_{xxxxx}=0\,.
\end{equation}

\subsection{$\mathcal{PT}$-symmetric rational extensions of 
the $\mathcal{PT}$-regularized Calogero systems}

The shape-invariant family  (\ref{HnPT})
is not a unique set of perfectly invisible
$\mathcal{PT}$-symmetric systems that
can be constructed by applying Darboux transformations to
the free particle.
A simple example of such a system of the form
different
from (\ref{HnPT}) can be obtained by taking
a wave function
\begin{equation}\label{psi*}
\psi_{\alpha,\gamma}^{(1)}={\gamma}{\xi^{-1}}+\xi^2
\end{equation}
as  a seed state for the Darboux transformation.
This is a  linear combination
of the bound state $\xi^{-1}$ of zero energy of
the system $H^{\alpha}_1$
and of its non-physical partner
in the sense  of
 (\ref{psitilde}),
where $\gamma$ is a constant.
Requiring that $\gamma$ is purely
imaginary,  the seed state (\ref{psi*}) will
be $\mathcal{PT}$-invariant.
Via a $\mathcal{PT}$-odd superpotential
\[
\mathcal{W}^{(1)}_{\alpha,\gamma}=-\frac{d}{dx}\left(\ln 
\psi_{\alpha,\gamma}^{(1)}\right)=\frac{1}{\xi}-\frac{3\xi^2}{\xi^3+\gamma}\,,
\]
one can generate  two super-partner potentials
$V_\pm=(\mathcal{W}^{(1)}_{\alpha,\gamma})^2\pm
(\mathcal{W}^{(1)}_{\alpha,\gamma})'$,
where $V_-=2\xi^{-2}$ and
\begin{equation}\label{U+}
V_+^{(1)}(x;\alpha,\gamma)=
6\xi\frac{\xi^3-2\gamma}{(\xi^3+\gamma)^2}=
\frac{6}{\xi^2}-6\gamma\frac{4\xi^3+\gamma}{\xi^2(\xi^3+\gamma)^2}\,.
\end{equation}
Potential (\ref{U+}) is a stationary $\mathcal{PT}$-symmetric
solution to the equation (\ref{KdV2}), which is non-singular 
on $\R^1$ function provided
\be\label{nuneq}
\gamma=i\nu \alpha^3\,,\qquad
\nu\in \R^1\,,\,\,\,\,\nu\neq -8\,,1.
\ee

On the other hand, if we
put  in (\ref{U+})
\be\label{nuneqt}
\gamma=\gamma(t)=12t+i\nu\alpha^3\,,\qquad
\nu\in(1,\infty)\,,
\ee
we obtain a complex $\mathcal{PT}$-symmetric
solution $V_+^{(1)}(x,t;\alpha)$ of the KdV equation (\ref{KdV}),
which is non-singular for $t\in(-\infty,\infty)$.
If in this time-dependent solution we put
$\alpha=0$, it takes the form of  the well known rational
singular
solution $u(x,t)=6x\frac{x^3-24t}{(x^3+12t)^2}$
to the KdV equation (\ref{KdV}) \cite{Draz}.
The permitted values of the parameter $\gamma$ 
indicated  in 
(\ref{nuneq}) and (\ref{nuneqt}) guarantee that 
the amplitude  of the wave function 
(\ref{psi*}) nowhere turns into zero.
Note that the fact that the potential
$V_+^{(1)}(x,t;\alpha)$ satisfies simultaneously 
the time-dependent KdV equation  (\ref{KdV})
and the higher order ordinary differential equation 
$30u^2 u_x -20 u_x u_{xx}-10uu_{xxx}+u_{xxxxx}=0$
being  a stationary case of (\ref{KdV2})
corresponds to a general property
mentioned in Section \ref{SecIntro} in relation 
to the Novikov equation.

\begin{figure}[htbp]
\begin{center}
\includegraphics[scale=0.6]{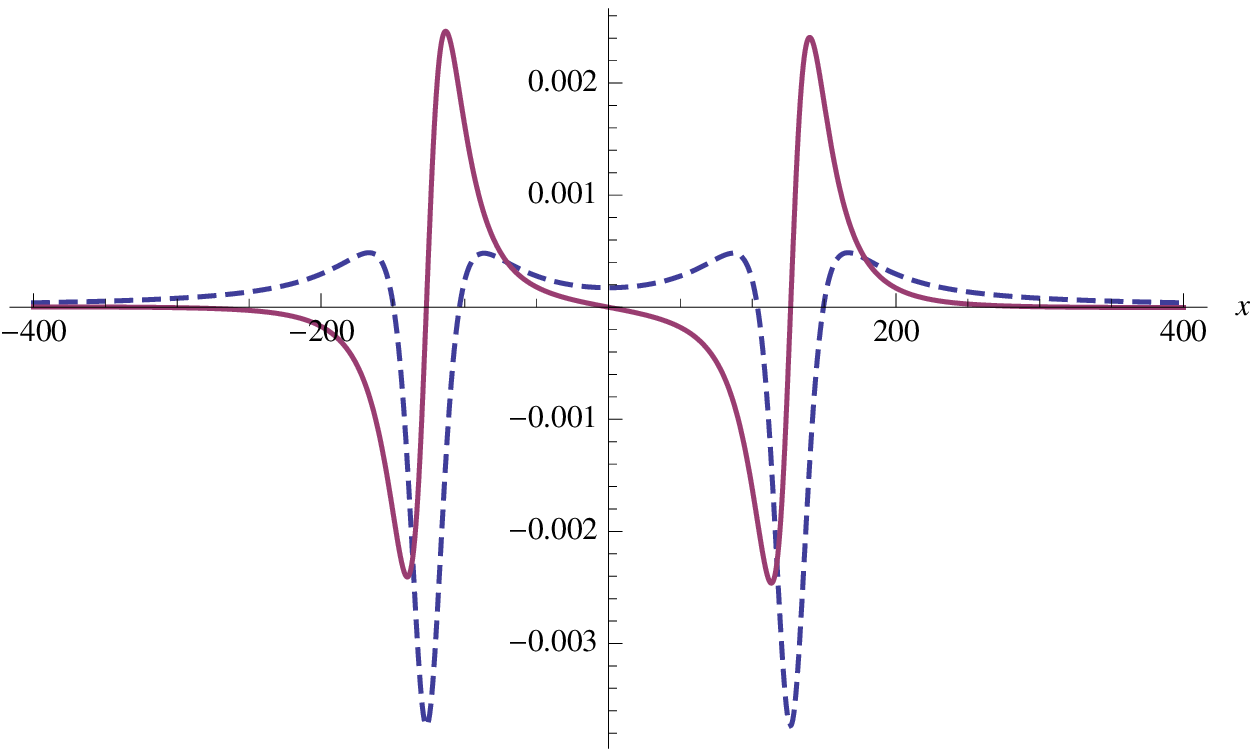}\includegraphics[scale=0.6]{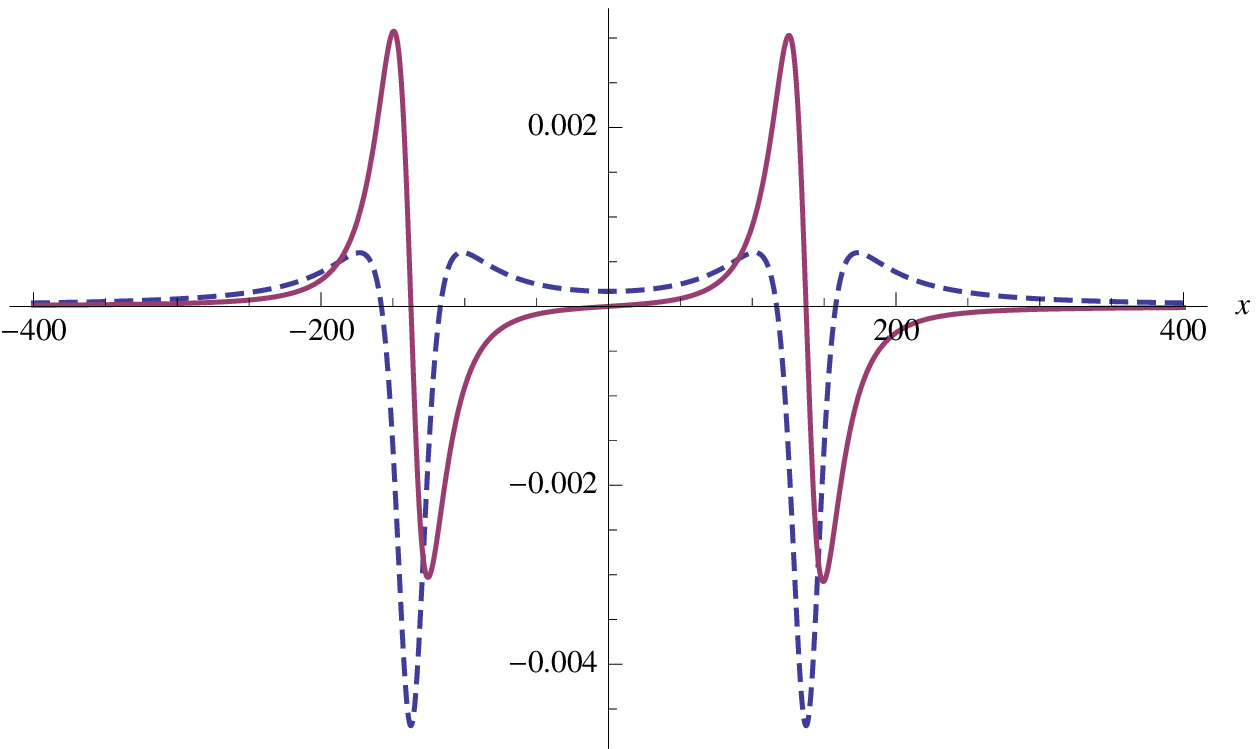}
\caption{  Real (shown by dashed lines) and imaginary (shown by continuous lines)  parts
of potential  (\ref{U+})
as a complex $\mathcal{PT}$-symmetric stationary ($t=0$) solution  
to  Eq. (\ref{KdV2}) at $\alpha=50$ and $\nu=-25$ (on the left) and 
$\alpha=100$, $\nu=-4$ (on the right). The graph of  imaginary 
part undergoes a flip as the parameter $\nu$ passes through a critical value $-8$.
}\label{FigV1ab}
\end{center}
\end{figure}

\begin{figure}[htbp]
\begin{center}
\includegraphics[scale=0.6]{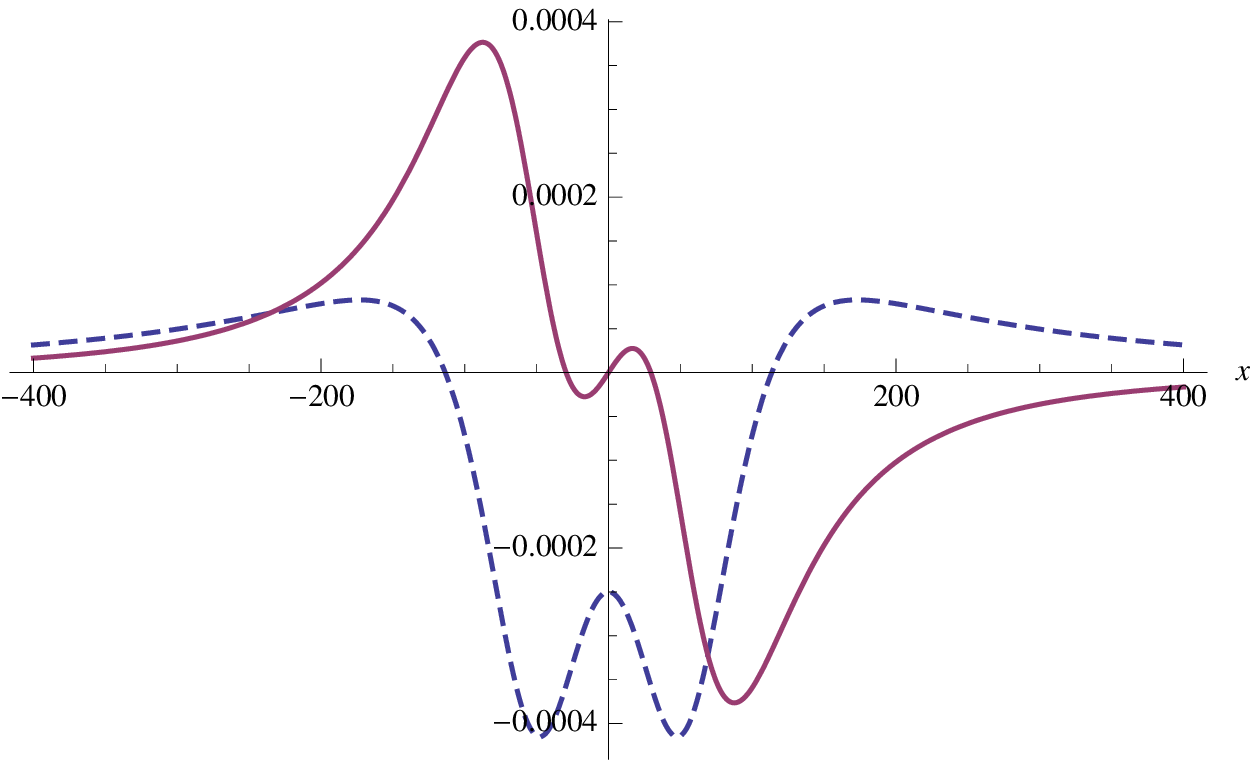}\includegraphics[scale=0.6]{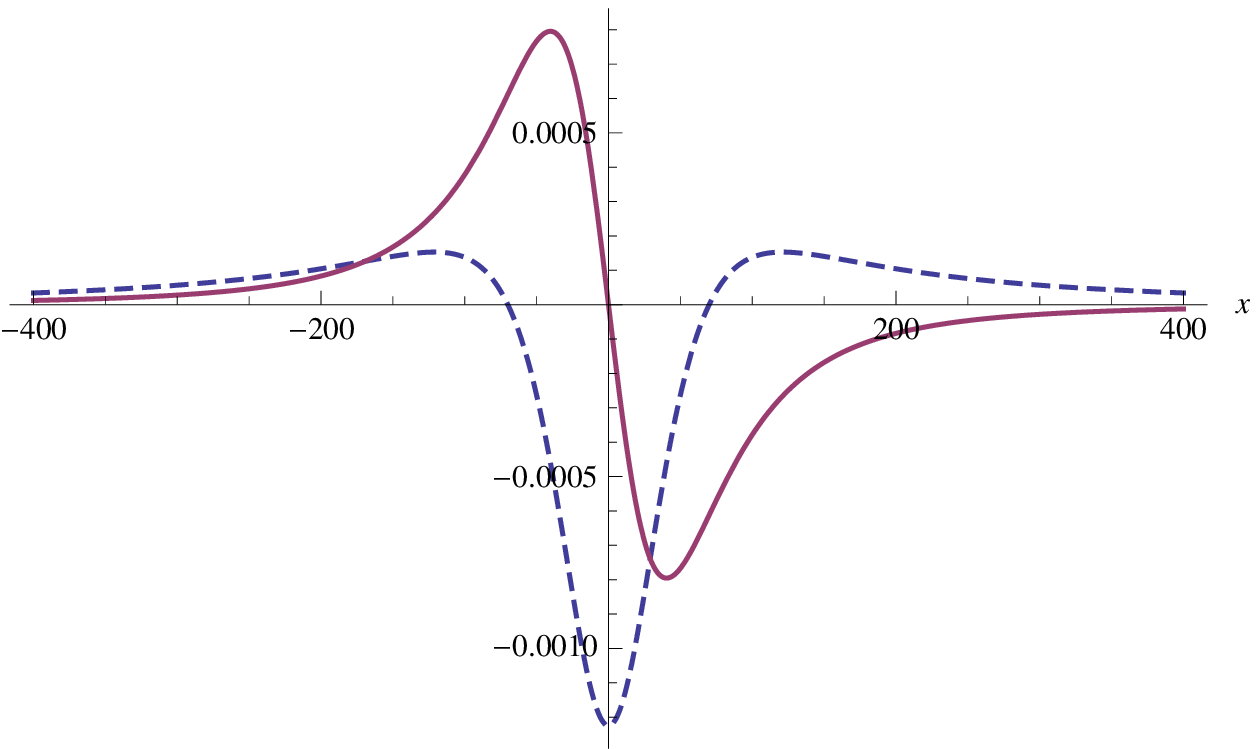}
\caption{Real  and imaginary parts
of (\ref{U+}) of a stationary case  (\ref{nuneq}) at $\alpha=100$, 
and $\nu=-0.2$ (on the left) and 
$\nu=0$ (on the right).
}\label{FigV1cd}
\end{center}
\end{figure}

The variation of the form of real and imaginary parts of
the potential (\ref{U+}), 
(\ref{nuneq})  as a stationary 
$\mathcal{PT}$-symmetric solution ($t=0$) 
to  the higher order KdV equation (\ref{KdV2}) are 
illustrated by Figures \ref{FigV1ab} and \ref{FigV1cd}
for
various  values of the parameters $\alpha$ and $\nu$.

\begin{figure}[htbp]
\begin{center}
\includegraphics[scale=0.6]{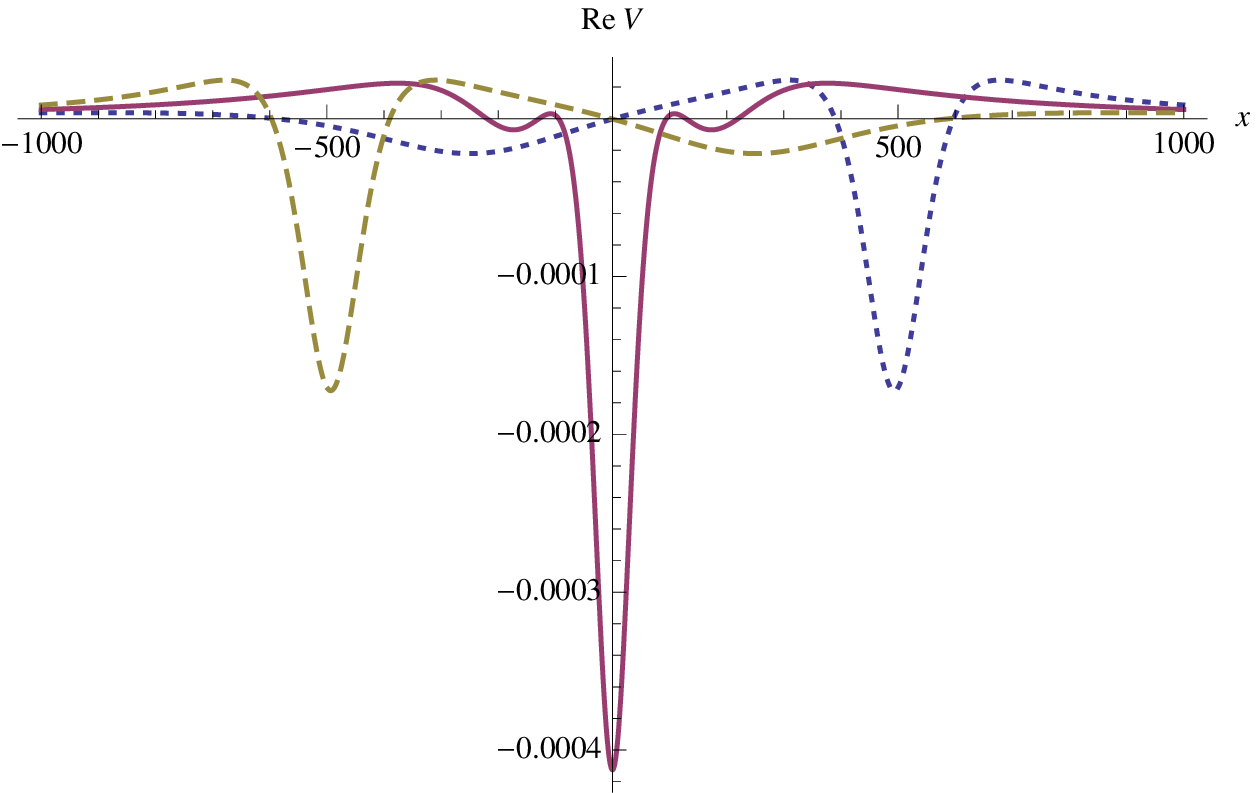}\includegraphics[scale=0.6]{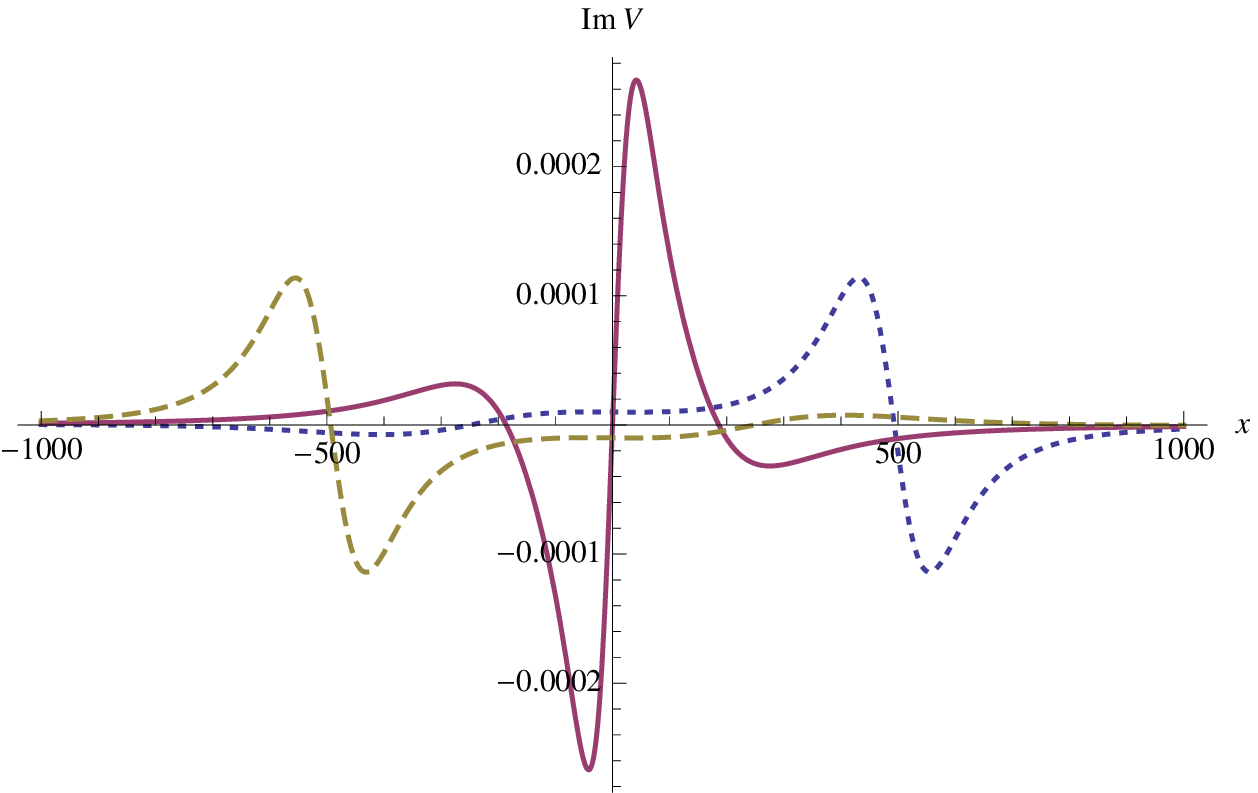}
\caption{  Evolution of real (on the left) and imaginary (on the right)  parts of
potential  (\ref{U+}), (\ref{nuneqt})
as a  complex $\mathcal{PT}$-symmetric solution 
of the KdV equation (\ref{KdV}) at $\alpha=100$, $\nu=5$;
dashed lines: $t=-10^{7}$, continuous lines: $t=0$, dotted lines: $t=10^{7}$.
}\label{FigV1Evolution}
\end{center}
\end{figure}

The evolution of this potential 
in the time-dependent case (\ref{nuneqt}) 
as a $\mathcal{PT}$-symmetric 
solution to the KdV equation (\ref{KdV}) is
shown on Figure \ref{FigV1Evolution}.
It is interesting to note that 
near the critical value $\nu=1$
of the parameter $\nu$, the graph 
of real part of the potential demonstrates  a singular 
$\delta$-function type behaviour
while the imaginary part 
undergoes a flip and
has a typical $\delta^\prime$
form.  
This critical value 
is at the very lower edge of the infinite 
interval (\ref{nuneqt})  of the permitted  values 
for  the parameter $\nu$ in the 
time-dependent case.
As a consequence, 
for the values of the parameter $\nu$ close to the 
critical value $\nu=1$,
the inverted  potential $-V_+^{(1)}(x;\alpha,\gamma(t))=u(x,t)$ 
reveals a behaviour typical for  \emph{rogue (extreme)} waves
that is seen already from  Fig. \ref{FigV1Evolution}.

On the base of the function (\ref{psi*}),
one can construct the first order
differential operators
\begin{equation}\label{D1psi*}
D^{(1)}_{\alpha,\gamma}=\psi^{(1)}_{\alpha,\gamma}
\frac{d}{dx}\frac{1}{\psi^{(1)}_{\alpha,\gamma}}=\frac{d}{dx}+\mathcal{W}^{(1)}_{\alpha,\gamma}\,,
\qquad
D^{(1)\#}_{\alpha,\gamma}=\frac{1}{\psi^{(1)}_{\alpha,\gamma}}\frac{d}{dx}\psi^{(1)}_{\alpha,\gamma}
=-\frac{d}{dx}+\mathcal{W}^{(1)}_{\alpha,\gamma}\,,
\end{equation}
which
 factorize  the pair of Hamiltonians
(\ref{H1PT}) and $H^{(1)}_{\alpha,\gamma}=-\frac{d^2}{dx^2}+V_+^{(1)}(x;\alpha,\gamma)$\,:
$D^{(1)\#}_{\alpha,\gamma} D^{(1)}_{\alpha,\gamma}=H^\alpha_1$,
$D^{(1)}_{\alpha,\gamma}D^{(1)\#}_{\alpha,\gamma} =H^{(1)}_{\alpha,\gamma}$\,,
and intertwine them\,:
$D^{(1)}_{\alpha,\gamma}H^\alpha_1=H^{(1)}_{\alpha,\gamma}
D^{(1)}_{\alpha,\gamma}$,
$D^{(1)\#}_{\alpha,\gamma}  H^{(1)}_{\alpha,\gamma}=
H^\alpha_1D^{(1)\#}_{\alpha,\gamma} $.
In correspondence with these relations,
the bounded eigenstates of the system $H^{(1)}_{\alpha,\gamma}$
are constructed from the plane wave eigenstates of the free particle
system,
$\psi^{(1)}_{\alpha,\gamma,\pm k}=D^{(1)}_{\alpha,\gamma}D_1e^{\pm ik\xi}$,
where the operator $D_1$ is given by Eq. (\ref{D1}).
The unique quadratically integrable eigenstate
$\psi^{(1)}_{\alpha,\gamma,0}(x)=\frac{\xi}{\xi^3+\gamma}$
of zero eigenvalue
of the system $H^{(1)}_{\alpha,\gamma}$  corresponds to the limit case
$k=0$ of the bounded  eigenstates
$\psi^{(1)}_{\alpha,\gamma,\pm k}$.
It
is generated  by applying the second order
operator $D^{(1)}_{\alpha,\gamma}D_1$
to the bounded eigenstate $\psi_{0,0}=1$ of the
free particle $H_0$,
or, equivalently, by applying the first order operator
$D^{(1)}_{\alpha,\gamma}$ to the ground state $\xi^{-1}$
of zero energy of $H^\alpha_1$.
This state is annihilated by the Lax-Novikov integral
\be
\mathcal{P}^{(1)}_{\alpha,\gamma}=-i
D^{(1)}_{\alpha,\gamma} D_1\frac{d}{dx}D_1^\#D^{(1)\#}_{\alpha,\gamma} \,,
\ee
$[H^{(1)}_{\alpha,\gamma},\mathcal{P}^{(1)}_{\alpha,\gamma}]=0$,
that satisfies the Burchnall-Chaundy relation
$\left(\mathcal{P}^{(1)}_{\alpha,\gamma}\right)^2=\left(H^{(1)}_{\alpha,\gamma}\right)^5$,
and distinguishes the bounded eigenstates
$\psi^{\alpha,\gamma}_{\pm k}$\,:
 $\mathcal{P}^{(1)}_{\alpha,\gamma} \psi^{\alpha,\gamma}_{\pm k}=
 \pm k^5\psi^{\alpha,\gamma}_{\pm k}$.
The potential (\ref{U+}) can be produced
directly from the free particle system
via the relation
\be\label{VnWn}
V_+^{(1)}(x;\alpha,\gamma)=-2\frac{d^2}{dx^2}\big(\ln W(\xi,-\gamma+\xi^3)\big)\,,
\ee
that corresponds to the second order Darboux-Crum
transformation.
Here the first argument of the Wronskian is
the non-physical eigenstate $\xi$ of zero eigenvalue
of $H_0$, whereas the second argument corresponds to a
linear combination of the  eigenstate $\psi_{0,0}=1$ and
the Jordan state $\xi^3$ of the second order, $(H_0)^2\xi^3=0$.

As yet another example of the $\mathcal{PT}$-symmetric
perfectly invisible
system we present here the system
described  by the potential
\be\label{V+10}
V_+^{(2)}(x;\alpha,\gamma)=\frac{2}{\xi^2}+
10\xi^3\frac{\xi^5-4\gamma}{(\xi^5+\gamma)^2}=
\frac{12}{\xi^2}-10\gamma\frac{6\xi^5+\gamma}{\xi^2(\xi^5+\gamma)^2}\,,
\ee
where $\gamma$ is, again, a purely imaginary  parameter.
This potential can be produced
from
the system $H_2^\alpha$ by constructing the Darboux transformation
based on the state
\be\label{psi2ag}
\psi^{(2)}_{\alpha,\gamma}=\gamma\xi^{-2}+\xi^3\,,
\ee
that is a linear combination of the quadratically integrable
zero energy eigenstate  $\xi^{-2}$ of $H_2^\alpha$ and of  its
non-physical partner $\xi^3$.
Equivalently, the perfectly transparent  potential (\ref{V+10})
can be produced by the relation of the form
(\ref{VnWn}) with the Wronskian changed for
$W(\xi,\xi^3,\frac{8}{3}\gamma+\xi^5)$,
where $\xi^5$ in the last argument
corresponds to the Jordan state of $H_0$ of the third order\,:
$(H_0)^3\xi^5=0$.
The potential (\ref{V+10}) is a stationary
$\mathcal{PT}$-symmetric solution to the equation
(\ref{KdV2}),
which is non-singular provided 
the following restriction on the values of the parameter $\gamma$ is introduced\,:
\be\label{nuneq+}
\gamma=i\nu \alpha^5\,,\qquad
\nu\in \R^1\,,\,\,\,\,\nu\neq -1\,,4,
\ee
under which the wave function (\ref{psi2ag}) nowhere turns into zero
on real line $x\in\R^1$.

The substitution 
\be\label{nuneqt+}
\gamma=\gamma(t)=-720t+i\nu\alpha^5\,,\qquad
\nu\in(24,\infty)\,,
\ee
in (\ref{V+10})
transforms the $\mathcal{PT}$-symmetric
stationary solution $V_+^{(2)}(x;\alpha,\gamma)$ of
(\ref{KdV2}) into  the $\mathcal{PT}$-symmetric function
 $V_+^{(2)}(x,t;\alpha)$ that is a
time-dependent $\mathcal{PT}$-symmetric solution of the
same equation (\ref{KdV2}) to be non-singular for 
all values of $t\in(-\infty,\infty)$. In the case $\gamma=0$, potential
(\ref{V+10}) reduces  to $\frac{12}{\xi^2}$,  that is a
potential of the $\mathcal{PT}$-symmetric perfectly
transparent system  (\ref{HnPT}) with $n=3$.

 The time dependence (\ref{nuneqt})
and (\ref{nuneqt+}) for the corresponding potentials 
considered here as well as in a  general case 
can be fixed by exploiting the covariance under Darboux transfromations of the
Lax representation for the KdV equation 
and for higher equations of the hierarchy mentioned 
in Section \ref{SecIntro}.

\begin{figure}[htbp]
\begin{center}
\includegraphics[scale=0.6]{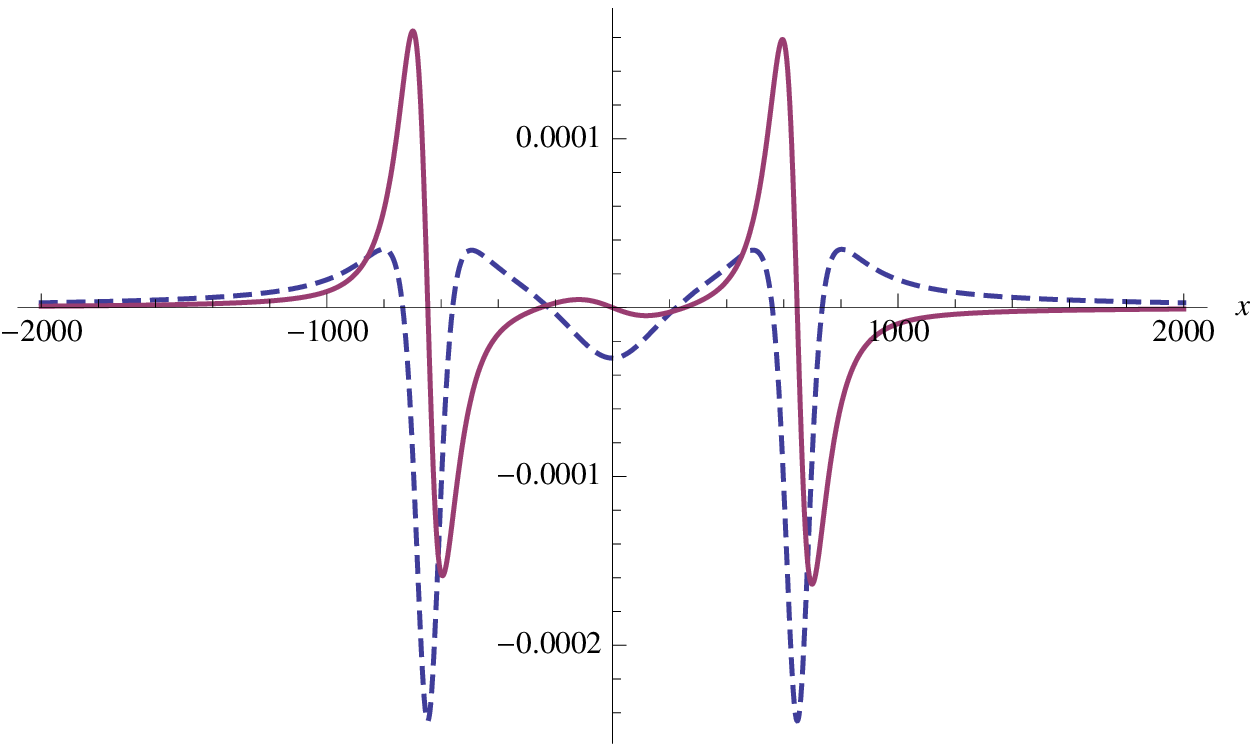}\includegraphics[scale=0.6]{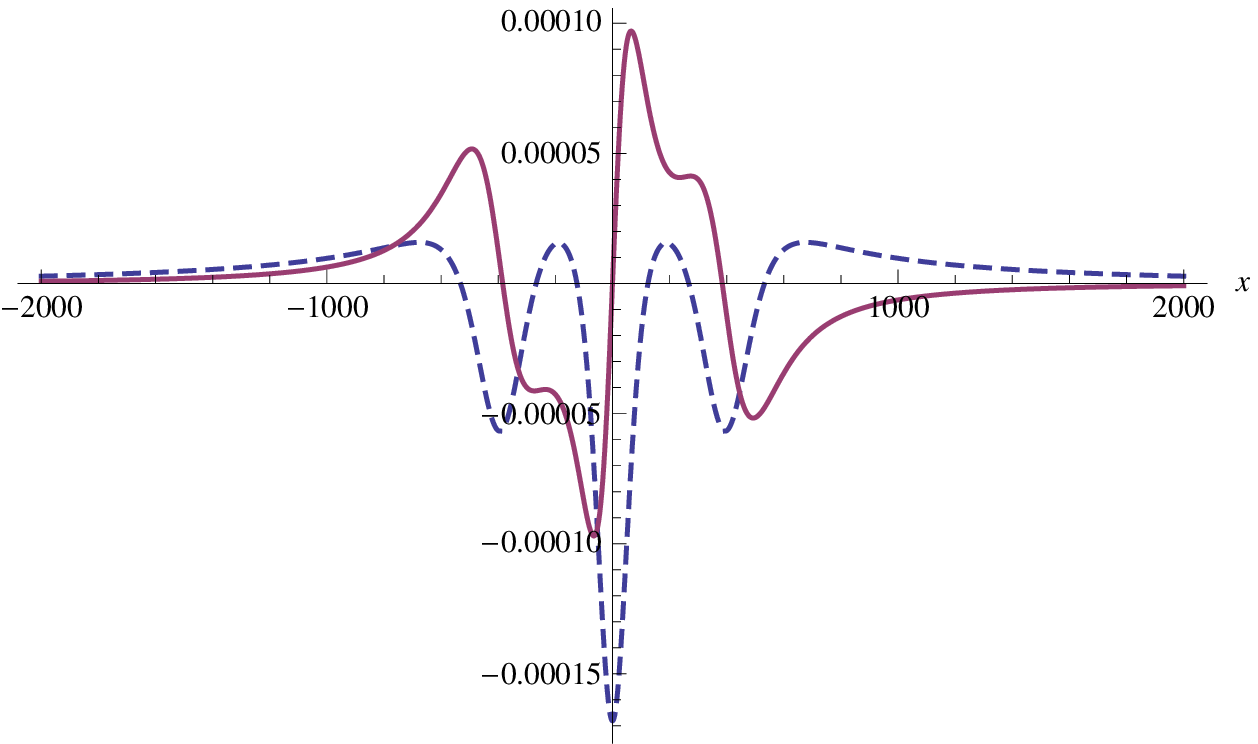}
\caption{  Real (shown by dashed lines) and imaginary (shown by continuous lines)  parts
of potential  (\ref{V+10}), (\ref{nuneq+})
as a complex $\mathcal{PT}$-symmetric stationary solution to Eq.
(\ref{KdV2}) at $\alpha=300$ and $\nu=-60$ (on the left) and 
$\nu=-5$ (on the right).
}\label{FigV2ab}
\end{center}
\end{figure}

\begin{figure}[htbp]
\begin{center}
\includegraphics[scale=0.6]{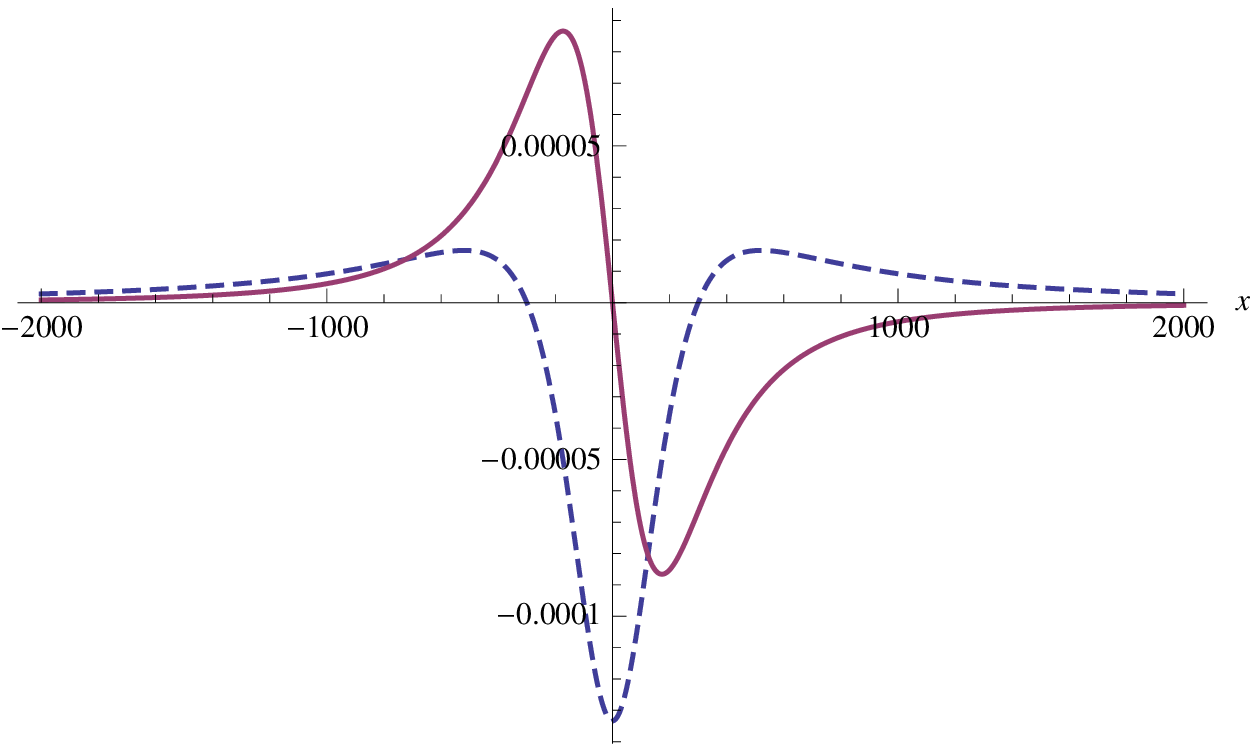}\includegraphics[scale=0.6]{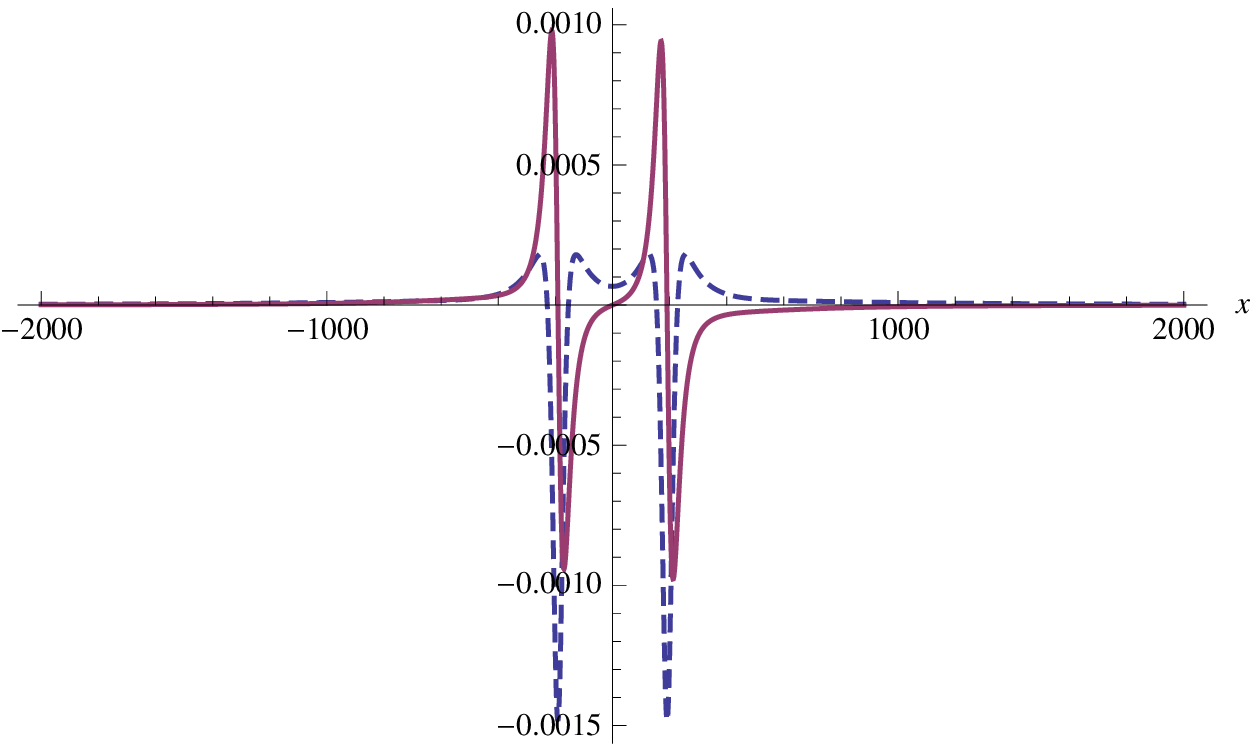}
\caption{  Real  and imaginary  parts
of potential (\ref{V+10}), (\ref{nuneq+})
as a complex $\mathcal{PT}$-symmetric stationary ($t=0$) solution  to Eq. 
(\ref{KdV2}) at $\alpha=300$ and $\nu=0$ (on the left) and 
$\nu=1.5$ (on the right).
}\label{FigV2cd}
\end{center}
\end{figure}

\begin{figure}[htbp]
\begin{center}
\includegraphics[scale=0.6]{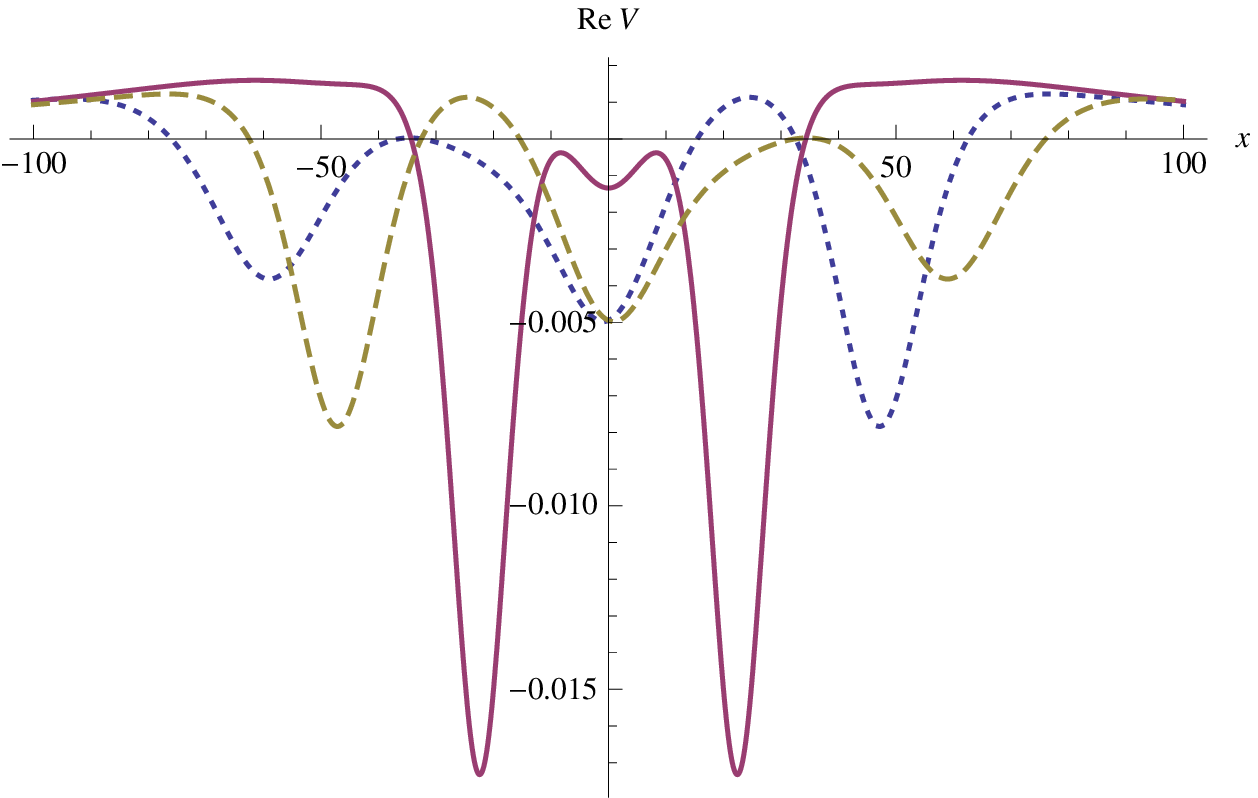}\includegraphics[scale=0.6]{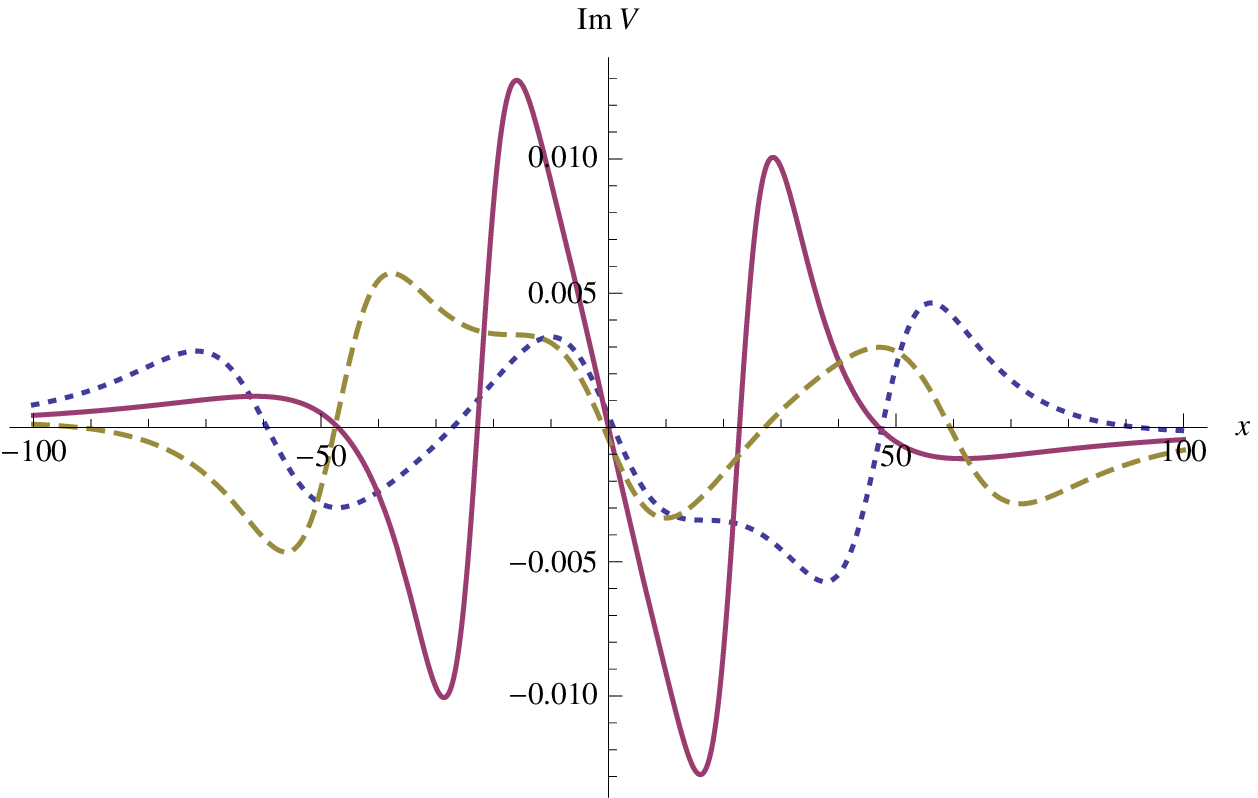}
\caption{  Evolution of  real (on the left) and imaginary (on the right)  parts of potential  (\ref{V+10}),
(\ref{nuneqt+})
as a  complex $\mathcal{PT}$-symmetric solution to Eq.  (\ref{V+10}) at $\alpha=20$, $\nu=25$;
dashed lines: $t=-10^{6}$, continuous lines: $t=0$, dotted lines: $t=10^{6}$.
}\label{FigV2Evolution}
\end{center}
\end{figure}

The form of potential (\ref{V+10})  
for  the stationary and time-dependent 
cases is illustrated by Figures \ref{FigV2ab}, \ref{FigV2cd} and
\ref{FigV2Evolution}.
Note that unlike the case corresponding to the
potential (\ref{U+}), the critical value $\nu=4$ 
here cannot be approached by the time-dependent solutions 
(\ref{nuneqt+}).


 The corresponding Hamiltonian operator
$H^{(2)}_{\alpha,\gamma}=-\frac{d^2}{dx^2}+V_+^{(2)}(x;\alpha,\gamma)$
is factorized,
$H^{(2)}_{\alpha,\gamma}=D^{(2)}_{\alpha,\gamma}D^{(2)\#}_{\alpha,\gamma}$,
and we also have $D^{(2)\#}_{\alpha,\gamma}D^{(2)}_{\alpha,\gamma}=H_2^\alpha$.
 Here the factorizing  operators
have the form similar to that in (\ref{D1psi*})
 with the generating function changed  for $\psi^{(2)}_{\alpha,\gamma}$,
 and with the superpotential replaced by
 \[
 \mathcal{W}^{(2)}_{\alpha,\gamma}=\frac{2}{\xi}
 -\frac{5\xi^4}{\gamma+\xi^5}\,.
 \]
The perfectly transparent $\mathcal{PT}$-symmetric
system $H^{(2)}_{\alpha,\gamma}$ is characterized
by the Lax-Novikov integral
\be\label{P2ag}
\mathcal{P}^{(2)}_{\alpha,\gamma}=-i
D^{(2)}_{\alpha,\gamma} D_2D_1\frac{d}{dx}D_1^\#D_2^\# D^{(2)\#}_{\alpha,\gamma} \,.
\ee
This  differential operator of order seven satisfies the relations
$[H^{(2)}_{\alpha,\gamma},\mathcal{P}^{(2)}_{\alpha,\gamma}]=0$
and
$(\mathcal{P}^{(2)}_{\alpha,\gamma})^2=(H^{(2)}_{\alpha,\gamma})^7$.
The  bounded eigenstates of the system $H^{(2)}_{\alpha,\gamma}$
are constructed from the plane wave eigenstates of the free particle
system by applying to them the third order differentia operator
$D^{(2)}_{\alpha,\gamma} D_2D_1$ that intertwines $H_0$ with
$H^{(2)}_{\alpha,\gamma}$.
The unique quadratically integrable eigenstate 
$\psi^{(2)}_{\alpha,\gamma,0}=\frac{\xi^2}{\xi^5+\gamma}$
of the system $H^{(2)}_{\alpha,\gamma}$ of zero eigenvalue
is given by application of the same third order
differential operator to the eigenstate $\psi_{0,0}=1$
of $H_0$. Equivalently,  $\psi^{(2)}_{\alpha,\gamma,0}$
 is produced by applying
the first order differential operator $D^{(2)}_{\alpha,\gamma} $
to the ground state $\xi^{-2}$ of zero energy of the
system $H^\alpha_2$.
It is annihilated by the Lax-Novikov integral
(\ref{P2ag}).

 Potential (\ref{U+})
 at $\gamma=0$ is reduced to the
 $\mathcal{PT}$-symmetric  potential
 $6\xi^{-2}$, while for $|\gamma|\rightarrow \infty$
 it turns into zero.
 Potential (\ref{V+10}) at $\gamma=0$ takes a form
 $12\xi^{-2}$, and in the limit
 $|\gamma|\rightarrow \infty$  it is reduced to
 $2\xi^{-2}$. Therefore, the
 quantum system with potential (\ref{U+})
 interpolates between the free particle and 
 the $\mathcal{PT}$-symmetric
 system  (\ref{HnPT})  with $n=2$,
 while the system with potential  (\ref{V+10})
 interpolates between the
 perfectly transparent  $\mathcal{PT}$-symmetric
 systems (\ref{HnPT})  with $n=1$ and $n=3$.
 Let us stress, however, that continuous  variation  of 
 $\gamma$ between $0$ and $\infty$  
 is impossible since,
 according to (\ref{nuneq}) and (\ref{nuneq+}), 
 we have to pass through the excluded purely imaginary  values of $\gamma$ 
 where the corresponding potentials loose their 
 non-singular in $x\in\R^1$  nature.

 It is interesting to note here what happens
 with the zero energy ground states
 of the systems $H^{(1)}_{\alpha,\gamma}$
 and $H^{(2)}_{\alpha,\gamma}$ in the indicated
 limit cases.
 The ground state
 $\psi^{(1)}_{\alpha,\gamma,0}=\frac{\xi}{\xi^3+\gamma}$
 of $H^{(1)}_{\alpha,\gamma}$ in the case
 $\gamma=0$ takes the form of the ground
 state $\xi^{-2}$ of $H^\alpha_2$,
 whereas the ground state
  $\psi^{(2)}_{\alpha,\gamma,0}=\frac{\xi^2}{\xi^5+\gamma}$
  of $H^{(2)}_{\alpha,\gamma}$
  at $\gamma=0$ reduces to the ground state
  $\xi^{-3}$ of $H^\alpha_3$.
 In the limit $|\gamma|\rightarrow \infty$,
 however,
 both states turn into zero.
 These states mutliplied by
 $\gamma$ in the indicated limit
 are transformed into the states
 $\xi$ and $\xi^2$, which are the
 zero energy non-physical eigenstates of
 $H_0$ and $H^\alpha_1$, respectively.
 On the other hand, the
 non-physical partner
 $\widetilde{\psi^{(1)}_{\alpha,\gamma,0}}$
 of the ground state $H^{(1)}_{\alpha,\gamma}$
 given by (\ref{psitilde}) and rescaled
 by inclusion of the multiplication factor
 $\gamma^{-1}$ 
 transforms in the  limit
 $|\gamma|\rightarrow \infty$  into the ground state
 $\psi_{0,0}=1$ of the system $H_0$.
 In the same way, the non-physical eigenstate
  $\gamma^{-1}\widetilde{\psi^{(2)}_{\alpha,\gamma,0}}$
  of $H^{(2)}_{\alpha,\gamma}$ of zero eigenvalue
  in the limit  $|\gamma|\rightarrow \infty$
  is transformed into the ground state
  $\xi^{-1}$ of the system $H^\alpha_1$.
  Therefore, in the limit  $|\gamma|\rightarrow \infty$,
  we have a kind of transmutation of physical and non-physical
  eigenstates of zero energy of the systems
  $H^{(1)}_{\alpha,\gamma}$ and
  $H^{(2)}_{\alpha,\gamma}$
  into, respectively, non-physical and physical states  of
  zero eigenvalue of the corresponding limit
  systems $H_0$ and $H^\alpha_1$.
  
In conclusion of this section 
 we  note  that the quantum systems given by the 
 $\mathcal{PT}$-symmetric potentials (\ref{U+}) and (\ref{V+10}), unlike 
 the $\mathcal{PT}$-regularized Calogero
 systems (\ref{HnPT}), are not conformal invariant.
 This can be related to the difference in the 
 properties  of the wave functions 
 (\ref{psi*}) and (\ref{psi2ag}) in comparison  
 with those of wave functions of the form 
 $\xi^{2n+1}$ with respect to the 
 scale, $\xi\rightarrow \lambda\xi$, and inversion,
 $\xi\rightarrow 1/\xi$,  transformations.

\section{$(1+1)$-dimensional conformal field theoretical kinks seen from conformal quantum mechanics}
\label{Section3}

In this Section we discuss how the Hamiltonians $H_1^\alpha$ and $H_2^\alpha$ 
governing the dynamics
in conformal invariant $\mathcal{PT}$-symmetric quantum mechanical systems also 
determine the fluctuation spectra around the singular
kinks arising as traveling waves in the Liouville and $SU(3)$ 
conformal Toda systems in field theory. This is remarkable
because the conformal group which is infinite dimensional in 
the two-dimensional Minkowskian setting contracts to a
finite subgroup in conformal quantum mechanics,
see also \cite{Vira1,Vira2}.

\subsection{Kinks in Liouville field theory}

The classical dynamics of Liouville field theory  \cite{JackiwdHoker,Jackiw} 
is governed by the action:
\begin{equation}
S[\phi]=\int \, dt\, \int \, dx \, 
\Big\{\frac{1}{2}\left(\frac{\partial\phi}{\partial t}\right)^2-
\frac{1}{2}\left(\frac{\partial\phi}{\partial x}\right)^2-
\frac{m^2}{\beta^2}\cdot \exp {\left[\beta \, \phi(t,x)\right]}\Big\} \, ,
\end{equation}
where $m^2$ is a parameter of dimension $({\rm length})^{-2}$ in the 
natural system of units $\hbar=c=1$. The coupling constant $\beta>0$, however, is dimensionless. 
The Liouville field equation
\begin{equation} \label{liouvpde}
\frac{\partial^2\phi}{\partial t^2}(t,x)-
\frac{\partial^2\phi}{\partial x^2}(t,x)+\frac{m^2}{\beta}\cdot \exp\left[\beta \, \phi(t,x)\right]=0
\end{equation}
is completely integrable. 
The general solution of (\ref{liouvpde}) is obtained, in terms of arbitrary functions of the light-cone coordinates
$x_\pm =\frac{1}{\sqrt{2}}(t\pm x)$,
via B\"acklund transformations starting from the general solution
of the free wave equation. In these variables the wave and Liouville
equations read respectively\,:
\begin{equation}
2\, \partial_+\partial_- \phi_0(x_+,x_-)=0 \,, 
\qquad  \qquad 2\, \partial_+\partial_-\phi(x_+,x_-)+
\frac{m^2}{\beta}\exp\,[\beta \phi(x_+,x_-)]=0 
\,,  \label{Liouc} 
\end{equation}
where we have denoted  
 $\partial_\pm=\frac{\partial}{\partial x_\pm}$. Equations (\ref{Liouc}) are thus invariant under 
 the coordinate transformations 
$x^\prime_+ =f_+(x_+)$,
$x^\prime_- =f_-(x_-)$, 
 with $f_\pm(x_\pm)$ arbitrary functions 
of their arguments, accomponied by the field redefinition 
$\phi^\prime(x^\prime_+,x^\prime_-)=-\frac{1}{\beta}\ln [ \partial_+f_+(x_+)\partial_- 
f_-(x_-)]+\phi(x_+,x_-)$ 
\cite{Jackiw}. In sum, 
the Liouville equation is invariant under the 
infinite-dimensional conformal group, whose Lie algebra is the Virasoro algebra.

There is a static homogeneous solution of equation 
(\ref{Liouc})\,: $\phi^0=-\infty$. Linearization of the Liouville equation around
the constant solution, $\phi(t,x)=\phi^0+\delta\phi(t,x)$,
\begin{equation}
\left(\frac{\partial^2}{\partial t^2}-\frac{\partial^2}{\partial x^2}\right)
\delta\phi(t,x)={\cal O}(\delta^2) \, \Rightarrow \, 
\delta\phi_\omega(t,x)=e^{i \omega t}e^{i k x}\, , 
\quad \omega^2=k^2\,,
\end{equation}
shows a semi-classical picture of the Liouville quanta as massless bosons. Liouville quantum theory in full 
exhibits  a very particular structure as a quantum conformal system, see e.g. \cite{JackiwdHoker}.
Our goal here, however, is the search for
traveling wave type of solutions,
that is, we make the ansatz $\phi(t,x)=f(\lambda\cdot\frac{x-v t-x_0}{\sqrt{1-v^2}})$, 
$\lambda\in\mathbb{R}^+$,
$\vert v\vert^2< 1$, $x_0\in\mathbb{R}$, which can be 
obtained from $\phi(t,x)= f(x)$ by applying  to $x$, first, 
a scale transformation, $x^\prime=\lambda x$, and then, 
a Poincar\'e transformation parametrized by $(v,x_0)$. 
Thus, in the frame $(\lambda=1, v=0,x_0=0)$,
one has to solve the first-order ordinary differential equation
\begin{equation}
\frac{d f}{dx}=\sqrt{2\frac{m^2}{\beta^2}\cdot \exp \left[\beta \, f(x)\right]} \, \, .
\end{equation}
The invariance under dilatations, Lorentz transformations 
and spatial translations provides the family of singular kinks
\begin{equation}
f_K(t,x)= \frac{1}{\beta} \log \left[\frac{2}{m^2\lambda^2(\frac{x-vt-x_0}{\sqrt{1-v^2}})^2}\right]\,,
\end{equation}
which have scales characterized by $\lambda^{-1}$, 
are singular at $x=v t+x_0$, and travel with velocity $-1<v<1$. The energy
\begin{equation}
E[f]= \int \, dx \, \left[\frac{1}{2}\left(\frac{df}{dx}\right)^2+U[f(x)]\right]
= \int \, dx \, \left[\frac{1}{2}\left(\frac{df}{dx}\right)^2+\frac{m^2}{\beta^2}
\cdot \exp\left[\beta \, f(x)\right]\right] \label{sen}
\end{equation}
of these singular solutions is ill-defined,
and   
computed at $x_0+v t=0$ 
is given by the improper integral
\begin{equation}
E[f_K]=\frac{1}{\lambda\beta^2} \int_{-\infty}^\infty \, dx \, \frac{4}{x^2} \, \, .
\end{equation}

Nevertheless, it is interesting to analyze the small fluctuation spectrum around the singular kink\,: 
$\phi(t,x)=f_K(x)+\delta \phi(t,x)$. The expansion of the action 
\begin{equation}
S[f_K+\delta\phi]-S[f_k]\simeq \frac{1}{2} \int\, dtdx \, \delta\phi (\lambda t, \lambda x)
\Big\{-\lambda^2\frac{d^2}{dx^2}+m^2{\rm exp}[\beta f_k(x)]\Big\}\delta\phi(\lambda t, 
\lambda x)+{\cal O}(\delta^3) \label{secorflu}
\end{equation}
around the singular kink shows that the second order fluctuations 
are governed by the differential operator 
\begin{equation}
K=-\frac{d^2}{dx^2}+\frac{d^2 U}{d \phi^2}\left(f_K(x)\right)=-\frac{d^2}{dx^2}+\frac{2}{x^2} \,,
 \label{conpot}
\end{equation}
which is precisely the Calogero Hamiltonian $H_1$ for two particles discussed in 
the previous Section. Moreover, the potential
in (\ref{conpot}) is a first 
 member of the family of Schr$\ddot{\rm o}$dinger 
operators appearing in de Alfaro-Fubini-Furlan 
quantum mechanics \cite{deAFF}.
The scale invariance $(t^\prime , x^\prime)= (\lambda t, \lambda x)$ manifest in the problem of
analyzing the kink fluctuations in Liouville conformal field theory, 
see (\ref{secorflu}), descends this way to 
a spectral problem in conformal quantum mechanics. 
We infer this link in the opposite direction as compared to the 
References \cite{Vira1,Vira2} where an ascending path 
 is travelled from conformal quantum mechanics to Virasoro algebra.
 
So far we have placed the kink singularity at the origin, $x_0+vt=0$, but
it is interesting to locate the kink center at the imaginary axis. 
Setting the scale to unity, $\lambda=1$, and 
putting the kink at rest, $v=0$, we choose also
 $x_0=-i\alpha$, $\alpha\in{\mathbb R}$, that is, 
we displace the kink center from $x_0=0$ to the imaginary 
axis in the complex $x$-plane as a way of escaping the singularity.  
Either at rest or in motion due to a Lorentz boost the kink 
profile with its center displaced to the 
imaginary axis has zero energy\,:
 \begin{equation}
 f_K[x,\alpha]=\frac{1}{\beta} \log \left[\frac{2(1-v^2)}{m^2 \lambda^2(x-vt+i\alpha)^2}\right] \, , 
 \quad E[f_K(\alpha]=\frac{1}{\lambda\beta^2}\int_{-\infty}^\infty \, dx \, \frac{4}{(x-vt+i\alpha)^2} =0 \, \, .
\end{equation}
Regarding small fluctuations around these displaced kinks, 
at rest $v=0$ 
with unit scale $\lambda=1$, 
which we  write in  the mod/arg form,
\[
f_K[x,\alpha]= \frac{1}{\beta} \log \left[\frac{2}{m^2(x^2+\alpha^2)}\right]+
\frac{i}{\beta}{\rm arg}[i\alpha +x]\, ,
\]
the second-order operator becomes\,:
\begin{equation}
K^\alpha = -\frac{d^2}{dx^2}+\frac{2}{(x+i\alpha)^2}=  -\frac{d^2}{dx^2}+ 
\frac{x^2-\alpha^2}{(x^2+\alpha^2)^2}-i\frac{\alpha x}{(
x^2+\alpha^2)^2}\, \, ,
\end{equation}
i.e., $K^\alpha=H_1^\alpha$ is the 
$\mathcal{PT}$-symmetric Hamiltonian discussed in Section \ref{Section2}.

In order to investigate the spectrum of Liouville kink fluctuations we profit from the
Darboux  transformation 
between $H_0=D_1^\dagger D_1$ and $H_1=D_1D_1^\dagger$. From the plane wave
odd eigenfunctions taming the singularity at the origin, in the spectrum of 
$H_0$, $H_0\sin kx=k^2\sin k x$, $k\in{\mathbb R}$, 
we obtain the eigenfunctions in the continuous spectrum of $H_1$\,:
\begin{equation}
H_1D_1(\sin k x)=D_1D_1^\dagger D_1(\sin k x)=
k^2\left(k  \cos k x- \frac{\sin k x}{x} \right).
\end{equation}
In addition, the kink fluctuations include a singular zero mode, $\delta\phi_0(x)=\frac{1}{x}$, 
which is not normalizable because of the singularity at the origin.
The kink fluctuations belonging to the continuous spectrum are
\[
 \delta\phi_\omega(t,x)=e^{i\omega t}\left(\frac{d}{dx}-\frac{1}{x}
 \right)\sin k x=e^{i \omega t}\left
 (k  \cos k x-\frac{\sin k x}{x}\right), \quad 
 \omega=k \,.
 \]
We again stress that the scale invariance of the Liouville field theory 
descends to scale invariance of the conformal quantum mechanics that governs 
the Liouville kink fluctuations. The dispersion
relation $\omega=k$ tells us that kink fluctuations 
are \lq\lq massless\rq\rq as a consequence of $(1+1)$-dimensional conformal symmetry.  
Comparing the $H_0$ and $H_1$ eigenfunctions 
at long distances, ${\rm sin}(k x)\propto e^{i k x}-e^{-ik x}$ 
versus  ${\rm cos}(k x)\propto e^{i k x}+e^{-ik x}$, we observe that a phase
$-1=e^{i \pi}$, $k$-independent, arises for the reflected waves in the $n=1$ Calogero 
potential non-existing in the free motion of one particle 
on the right half-line  with 
Dirichlet boundary conditions, see Appendix. This phase
can be interpreted from the well studied fermionic character of kinks and solitons.

In the case of the kinks displaced into the imaginary 
axis things are identical with respect to small fluctuations except that now the
Darboux-related operators
\begin{equation}
H_0=D_1^\#D_1=-\frac{d^2}{d x^2} \, \, \quad {\rm and} \, \,\quad H_1^\alpha=
D_1D_1^\#=-\frac{d^2}{dx^2}+\frac{2}{(x+i\alpha)^2}
\end{equation}
are intertwined by the first-order operators $D_1=\frac{d}{dx}-\frac{1}{x+i\alpha}$ 
and $D_1^\#=-\frac{d}{dx}-\frac{1}{x+i\alpha}$. 
Starting from the odd eigenfunctions of the free particle
we find through the action of $D_1$ the continuous spectrum eigenfunctions of $K_1^\alpha$\,:
\begin{equation}
\delta\phi_\omega(t,x)=e^{i\omega t}\left(\frac{d}{dx}-\frac{1}{x+ i\alpha}\right)\sin k x= e^{i\omega t} 
\left( k \cos k x- \frac{x-i\alpha}{x^2+\alpha^2}\,\sin k x \right)\,, \quad \omega=k \, .
\end{equation}
In this case the real part of these eigenfunctions is non-null at the origin, 
${\rm Re}\,\delta\phi_\omega(0,0)=k$, but the imaginary part vanishes at $(t=0,x=0)$:
 ${\rm Im}\,\delta\phi_\omega(0,0)=0$. Besides the continuous spectrum 
 eigenfunctions,
  there is a kink zero mode of fluctuation annihilated by $D_1^\#$,
\begin{equation}
\frac{d\delta\phi_0}{dx}+\frac{1}{x+i\alpha}\delta\phi_0(x)=0 
\quad \Rightarrow \quad \delta\phi_0(x)=\frac{1}{x+ i\alpha}\, ,
\end{equation}
which is regular and quadratically integrable on real line, 
but is purely imaginary at $x=0$.

\subsection{Kinks in the $SU(3)$ conformal Toda model}

The $SU(3)$ conformal Toda field theory \cite{BilalGervais} encompasses two $(1+1)$-dimensional 
scalar fields assembled into a vector 
field through use of the simple roots
of the $SU(3)$ Lie algebra: $\vec{\alpha}_1\cdot\vec{\alpha}_1=\vec{\alpha}_2\cdot\vec{\alpha}_2=2$ , 
 $\vec{\alpha}_1\cdot\vec{\alpha}_2=\vec{\alpha}_2\cdot\vec{\alpha}_1=-1$,
\[
\vec{\psi}(t,x)=\vec{\alpha}_1 \psi_1(t,x)+\vec{\alpha}_2\psi_2(t,x) \, , 
\qquad  C=2\left(\begin{array}{cc} \frac{\vec{\alpha}_1\cdot\vec{\alpha}_1}{\vert\vec{\alpha}_1
\vert\cdot\vert\vec{\alpha}_1\vert}
 & \frac{\vec{\alpha}_1\cdot\vec{\alpha}_2}{\vert\vec{\alpha}_1\vert\cdot\vert\vec{\alpha}_2\vert} \\ & 
  \\ \frac{\vec{\alpha}_2\cdot\vec{\alpha}_1}{\vert\vec{\alpha}_2\vert\cdot\vert\vec{\alpha}_1\vert} &
   \frac{\vec{\alpha}_2\cdot\vec{\alpha}_2}{\vert\vec{\alpha}_2\vert\cdot\vert\vec{\alpha}_2\vert}
   \end{array}\right)=\left(\begin{array}{cc} 2 & -1 \\ -1 
  & 2 \end{array}\right) \,,
  \]
where  $C$ is the $SU(3)$ Cartan matrix. The conformal Toda Lagrangian is
\begin{equation}
{\cal L}[\vec{\psi}]=\frac{1}{2}\frac{\partial\vec{\psi}}{\partial t}\cdot\frac{
\partial\vec{\psi}}{\partial t}-\frac{1}{2}\frac{\partial\vec{\psi}}{\partial x}\cdot\frac{
\partial\vec{\psi}}{\partial x}-\frac{m^2}{\beta^2}\Big[ \exp [\frac{\beta}{\sqrt{2}}\vec{\alpha}_1
\cdot\vec{\psi}(t,x)]+\exp [\frac{\beta}{\sqrt{2}}\vec{\alpha}_2\cdot\vec{\psi}(t,x)]\Big]\,,
\end{equation}
and the field equations read
\begin{eqnarray}
 4\, \frac{\partial^2\psi_1}{\partial x_+\partial x_-}-2\, \frac{\partial^2\psi_2}{\partial x_+
 \partial x_-}+\frac{m^2}{\beta}\Big[2\, \exp [\frac{\beta}{\sqrt{2}}(2\psi_1-\psi_2)]-
 \exp [\frac{\beta}{\sqrt{2}}(2\psi_2-\psi_1)]\Big]=0\,, \label{toda1} &&\\  4\, 
 \frac{\partial^2\psi_2}{\partial x_+\partial x_-}-2\, \frac{\partial^2\psi_1}{\partial x_+
 \partial x_-}+\frac{m^2}{\beta}\Big[2\, \exp [\frac{\beta}{\sqrt{2}}(2\psi_2-\psi_1)]-
 \exp [\frac{\beta}{\sqrt{2}}(2\psi_1-\psi_2)]\Big]=0
  \label{toda2}  \, . &&
\end{eqnarray}
The PDE system (\ref{toda1}), (\ref{toda2}) is manifestly conformal invariant if the
 space-time transformations $x^\prime_\pm=f_\pm(x_\pm)$ are accompanied by 
 the field redefinitions\,:
\[
\psi_1^\prime(x^\prime_+,x^\prime_-)=-\frac{\sqrt{2}}{\beta}\ln [ \partial_+f_+
(x_+)\partial_- f_-(x_-)]+\psi_1(x_+,x_-) \,,
\]
\[
\psi_2^\prime(x^\prime_+,x^\prime_-)=-\frac{\sqrt{2}}{\beta}\ln [ \partial_+
f_+(x_+)\partial_- f_-(x_-)]+\psi_2(x_+,x_-) \,.
\]
Since the eigenvectors of the Cartan matrix $C$ are $v_1=\left(\begin{array}{c} 1 \\ 1
 \end{array}\right)$ and $v_2=\left(\begin{array}{c} 1 \\ -1 \end{array}\right)$, 
with eigenvalues $1$ and $3$,  respectively, it is convenient to perform the following 
 change of variables in the fields\,: $\phi_1=\frac{1}{\sqrt{2}}(\psi_1+\frac{1}{\sqrt{3}} \psi_2)$ 
 and $\phi_2=\frac{1}{\sqrt{2}}(\psi_1-\frac{1}{\sqrt{3}} \psi_2)$.
In the new fields the $SU(3)$ 
conformal Toda field theory action becomes\,:
\begin{eqnarray}
S[\phi_1,\phi_2]&=&\int\int dt\, dx \, \Big\{\frac{1}{2}\left(\frac{\partial\phi_1}{\partial t}\right)^2+
\frac{1}{2}\left(\frac{\partial\phi_2}{\partial t}\right)^2-\frac{1}{2}\left(\frac{\partial\phi_1}{\partial x}\right)^2
-\frac{1}{2}\left(\frac{\partial\phi_2}{\partial x}\right)^2- U(\phi_1,\phi_2)\Big\}\,,\nonumber\\
U(\phi_1,\phi_2)&=&\frac{m^2}{\beta^2}\cdot\Big\{  \exp \left[\frac{\beta}{2}(\phi_1(t,x)+\sqrt{3}\, 
\phi_2(t,x))\right]+\exp \left[\frac{\beta}{2}(\phi_1(t,x)-\sqrt{3} \, \phi_2(t,x))\right]\Big\}\, \, , \nonumber
\end{eqnarray}
and the Toda field equations are diagonal in the field derivatives\,:
\begin{eqnarray}
&& \hspace{-1cm}\frac{\partial^2\phi_1}{\partial t^2}(t,x)-\frac{\partial^2\phi_1}{\partial x^2}(t,x)+
\frac{m^2}{2\beta}\, \left(e^{\frac{\beta}{2}(\phi_1(t,x)+\sqrt{3} \, 
\phi_2(t,x))}+e^{\frac{\beta}{2} \, 
(\phi_1(t,x)-\sqrt{3}\, \phi_2(t,x))}\right)=0\,,
 \label{todfe1}\\ && \hspace{-1cm}\frac{\partial^2\phi_2}{\partial t^2}(t,x)-
\frac{\partial^2\phi_2}{\partial x^2}(t,x)+\frac{\sqrt{3} m^2}{2\beta}\, \left(e^{\frac{\beta}{2}(\phi_1(t,x)+
\sqrt{3} \, \phi_2(t,x))}-e^{\frac{\beta}{2} \, (\phi_1(t,x)-\sqrt{3}\, \phi_2(t,x))}\right)=0 \,. \label{todfe2} 
\end{eqnarray}
Like the Liouville PDE this system of PDE's is completely integrable in terms of arbitrary 
functions of $x_+$ and $x_-$ with the help of the infinite dimensional conformal symmetry
with Lie algebra being some variant of Kac-Moody algebra.

The second-order variations around any solution $(\phi_1^S(t,x),\phi_2^S(t,x))$ of these 
PDE system are governed by the Hessian operator
\begin{equation}
{\cal H}=\left(\begin{array}{cc} \frac{\partial^2}{\partial t^2}-\frac{\partial^2}{\partial x^2}+ 
\frac{\delta^2 U}{\delta\phi_1^2}(\phi_1^S(t,x),\phi_2^S(t,x)) &   \frac{\delta^2 U}{\delta\phi_1
\delta \phi_2}(\phi_1^S(t,x),\phi_2^S(t,x)) \\ \frac{\delta^2 U}{\delta\phi_2 \delta\phi_1}(
\phi_1^S(t,x),\phi_2^S(t,x)) & \frac{\partial^2}{\partial t^2}-\frac{\partial^2}{\partial x^2}+
\frac{\delta^2 U}{\delta\phi_2^2}(\phi_1^S(t,x),\phi_2^S(t,x)) \end{array}\right),
\end{equation}
where
\begin{eqnarray}
\frac{\delta^2 U}{\delta\phi_1^2}(\phi_1^S(t,x),\phi_2^S(t,x))&=&\frac{m^2}{4}
\left[e^{\frac{\beta}{2}(\phi_1^S(t,x)+\sqrt{3} \phi_2^S(t,x))}+e^{\frac{\beta}{2} \, (
\phi_1^S(t,x)-\sqrt{3} \phi_2^S(t,x))}\right],\\
\frac{\delta^2 U}{\delta\phi_1\delta\phi_2}(\phi_1^S(t,x),\phi_2^S(t,x))&=&\frac{\sqrt{3} 
\,m^2}{4}\left[e^{\frac{\beta}{2}(\phi_1^S(t,x)+\sqrt{3} \phi_2^S(t,x))}-e^{\frac{\beta}{2} \, 
(\phi_1^S(t,x)-\sqrt{3}\phi_2^S(t,x))}\right], \\ \frac{\delta^2 U}{\delta\phi_2\delta\phi_1}
(\phi_1^S(t,x),\phi_2^S(t,x))&=&\frac{\sqrt{3}\, m^2}{4}\left[e^{\frac{\beta}{2}(\phi_1^S(t,x)+
\sqrt{3} \phi_2^S(t,x))}-e^{\frac{\beta}{2} \, (\phi_1^S(t,x)-\sqrt{3}\phi_2^S(t,x))}\right], \\ \frac{\delta^2 U}{
\delta\phi_2^2}(\phi_1^S(t,x),
\phi_2^S(t,x))&=&\frac{3 m^2}{4}\left[e^{\frac{\beta}{2}(\phi_1^S(t,x)+\sqrt{3} \phi_2^S(t,x))}+
e^{\frac{\beta}{2} \, (\phi_1^S(t,x)-\sqrt{3} \phi_2^S(t,x))}\right].
\end{eqnarray}
We first consider the only solution of the PDE system (\ref{todfe1}), (\ref{todfe2}) 
independent of $t$ and $x$\,: $(\phi_1^0
=-\infty, \phi_2^0=0)$, 
which is the 
absolute minimum of $U(\phi_1,\phi_2)$. Clearly, the Hessian evaluated at this 
constant solution is 
${\cal H}^0=\left(\begin{array}{cc} \frac{\partial^2}{\partial t^2}-\frac{\partial^2}{\partial x^2} &  
 0 \\ 0 & \frac{\partial^2}{\partial t^2}-\frac{\partial^2}{\partial x^2} \end{array}\right)$. 
 Since the spectrum of ${\cal H}^0$
 is positive,  the solution $(\phi_1^0,\phi_2^0)$ is classicaly stable with respect to small fluctuations. 
 Thus, in the quantum framework these fluctuations of ``vacuum" become
  massless bosons (compatible with conformal symmetry).

The next step is the search for solitary wave type of solutions,
that is, making the ansatz $(\phi_1(t,x)=f(x), \phi_2(t,x)=0)$ (in the kink center of mass), 
the PDE system above reduces to the first-order ODE
\begin{equation}
\frac{d f}{dx}=2\frac{m}{\beta}\cdot \exp\left[\frac{\beta}{4} \, f(x)\right] \, .
\end{equation}
Setting the kink scale to unit, $\lambda=1$,
Lorentz and translation invariance provide the family of singular kinks
\begin{equation}
f_K(t,x)= \frac{2}{\beta} \log \left[\frac{4(1-v^2)}{m^2(x-vt-x_0)}\right] \label{tkl}
\end{equation}
traveling  with velocity $-1<v<1$. The energy $E[(f,0)]= \int \, dx \, \left[\frac{1}{2}\left(\frac{df}{dx}\right)^2+
2\frac{m^2}{\beta^2}\cdot\exp \left[\frac{\beta}{2} \, f(x)\right]\right]$ of these singular solutions
is ill defined. Computed at $x_0=0$ it is given by the improper integral
\begin{equation}
E[(f_K,0)]=\frac{1}{\beta^2}\int_{-\infty}^\infty \, dx \, \frac{4}{x^2} \, \, .
\end{equation}
Like in the case of the Liouville kinks it is interesting to analyze the small 
fluctuation spectrum around these Toda singular kinks\,: 
$(\phi_1(t,x)=f_K(x)+
\delta \phi_1(t,x),\phi_2(t,x)=\delta\phi_2(t,x))$. 
The fluctuations up to second-order 
${\cal O}(\delta^2)$ belong to the kernel of 
${\cal H}(f_K(x),0)$\,:
\begin{equation}
{\cal H}(f_K(x),0)=\left(\begin{array}{cc} 
\frac{\partial^2}{\partial t^2} & 0 \\ 0 &
 \frac{\partial^2}{\partial t^2}\end{array}\right)+K =
 \left(\begin{array}{cc} \frac{\partial^2}{\partial t^2} & 
 0 \\ 0 & \frac{\partial^2}{\partial t^2}\end{array}\right)+
 \left(\begin{array}{cc}-\frac{\partial^2}{\partial x^2}+
 \frac{2}{x^2} & 0 \\ 0 & -\frac{\partial^2}{\partial x^2}+
 \frac{6}{x^2}  \end{array}\right). \label{hess}
\end{equation}
The Toda kink fluctuations thus can be expanded in terms of the eigenfunctions of the matrix differential 
operator $K$\,:
\begin{equation}
K\left(\begin{array}{c} \delta\phi_1^\omega(x) \\ \delta\phi_2^\omega(x)\end{array}\right)=
\left(\begin{array}{cc}-\frac{d^2}{dx^2}+\frac{2}{x^2} & 0 \\ 0 & -\frac{d^2}{d x^2}+\frac{6}{x^2}
  \end{array}\right)\left(\begin{array}{c} \delta\phi_1^\omega(x) \\ \delta\phi_2^\omega(x)
  \end{array}\right)=\omega^2\left(\begin{array}{c} \delta\phi_1^\omega(x) \label{operK}\\ 
  \delta\phi_2^\omega(x)\end{array}\right) .
  \end{equation}
{}From the explicit form of the hessian (\ref{hess}) we clearly see that its kernel ${\cal H}\delta\phi(t,x)$
is Fourier expanded in terms of the ${\rm K}$-eigenfunctions: 
\begin{equation}
{\cal H}\delta\phi(t,x)=0 \, \, \, \Leftrightarrow \, \, \, \delta\phi(t,x)=\int_{-\infty}^\infty \,d\omega\, 
e^{i \omega t}\left(\begin{array}{c} \delta\phi_1^\omega(x) \\ \delta\phi_2^\omega(x)
\end{array}\right) .
\end{equation}
Since $K$ is a diagonal matrix of two-particle Calogero Hamiltonians,
the conformal Toda kink fluctuations exhibit similar features
to the properties of Liouville kink fluctuations just analyzed in the previous sub-Section. 
In particular, the fluctuations in the
$\phi_1$ field, $\delta\phi_1^\omega(x)$, are exactly the eigenfunctions in the continuous spectrum
 of $H_1$ vanishing at $x=0$. As for Liouville kinks it is possible to escape from the singularity 
 at $x=0$ by sending the kink center to the imaginary axis.

The novelties arise in the $\delta\phi_2(t,x)$ fluctuations. These fluctuations may be
 expanded in terms of the eigenfunctions of
$H_2$, the $n=2$ two-particle Calogero Hamiltonian $H_2$\,:
\begin{equation}
\delta\phi_2^\omega(x)=\left(\frac{d}{dx}-\frac{2}{x}\right)\left(\frac{d}{dx}-
\frac{1}{x}\right) \sin k x=-\frac{3 k}{x}\cos k x
+ \frac{3-k^2 x^2}{x^2}\sin k x \,, \quad  \omega^2=k^2 \,.
\end{equation}
This is possible because the hierarchy
\begin{eqnarray}
H_0=D_1^\dagger D_1\, , \qquad H_1&=& D_1 D_1^\dagger\,, \\ 
H_1&=& 
D_2^\dagger D_2 \,, \qquad H_2= D_2 D_2^\dagger
\end{eqnarray}
allows us to obtain the eigenfunctions of $H_2$ iteratively from the $H_0$ 
eigenfunctions\,:
\begin{equation}
H_2\left(-\frac{3 k}{x}\cos k x
+ \frac{3-k^2 x^2}{x^2}\sin k x\right)=k^2 \left(-\frac{3 k}{x} \cos k x
+ \frac{3-k^2 x^2}{x^2}\sin k x \right) \,.
\end{equation}
Besides the positive eigenfunctions, the spectrum of $H_2$ includes a singular zero mode\,:
\begin{equation}
D_2^\dagger \delta\phi_2^0 (x)=0 \quad \Rightarrow \quad \delta\phi_2^0 (x)\propto \frac{1}{x^2} \, \, ,
\end{equation}
to be added to the kink translational mode: $\delta\phi_1^0 (x)\propto \frac{1}{x} $.

We omit the description of these singular Toda kinks regularized by pushing 
their centers to the imaginary axis because their
properties are essentially identical to those of regular Liouville kinks. 
We conclude this Section by mentioning that the one-loop
mass shifts of Liouville and Toda kinks are null. Since the kink Hessians 
are reflectionless,  the Cahill-Comtet-Glauber formula 
\cite{CCG,JMGPS,AlAIJuMG} can be
applied. Because the discrete and continuous spectra collapse at 
the point $k=0$, the threshold of the continuous spectrum, the CCG formula gives
mass shifts equal to zero. This is because the CCG formula expresses the one-loop kink mass shift as 
a sum of terms collecting the bound state contributions times the scattering threshold that in this case is zero. 
The annihilation of one-loop kink mass shifts
is thus the consequence of perfect invisibility of the Calogero Hamiltonians.

 \section{Exotic supersymmetry of the pairs of
 perfectly invisible $\mathcal{PT}$-symmetric
 systems}\label{Section4}

In this section we investigate the exotic nonlinear supersymmetry 
structure appearing in  the extended systems composed from the pairs 
of  perfectly invisible $\mathcal{PT}$-symmetric 
quantum systems related by the first order Darboux transformations.

Consider the simplest extended system composed
from the pair $H_0$ and $H^\alpha_1$
and described by the matrix Hamiltonian
\be\label{Hextend}
\mathcal{H}=
\left(
\begin{array}{cc}
 H^\alpha_1 & 0   \\
0  &  H_0
\end{array}
\right).
\ee
The extended system has the integrals
of motion
\be\label{Qa}
Q_1=
\left(
\begin{array}{cc}
 0 & D_1   \\
D^\#_1  &  0
\end{array}
\right),\qquad
Q_2=i\sigma_3Q_1\,,
\ee
\be\label{Sa}
S_1=
\left(
\begin{array}{cc}
 0 & -iD_1 \mathcal{P}_0  \\
i\mathcal{P}_0 D^\#_1  &  0
\end{array}
\right),\qquad
S_2=i\sigma_3S_1\,,
\ee
where $D_1$ is the first order differential operator
given by Eq.  (\ref{D1}),
and  $\mathcal{P}_0=-i\frac{d}{dx}$ is the
momentum operator of the free particle $H_0$.
Operators $Q_a$ and $S_a$, $a=1,2$,
constitute the set of four supercharges
of the extended system\,: they commute with
the Hamiltonian $\mathcal{H}$,
\be\label{susyHQS}
[\mathcal{H},Q_a]=0\,,\qquad
[\mathcal{H},S_a]=0\,,
\ee
 and their anticommutation
relations are
\be\label{susyQQSS}
\{Q_a,Q_b\}=2\delta_{ab}\mathcal{H}\,,\qquad
\{S_a,S_b\}=2\delta_{ab}\mathcal{H}^2\,,
\ee
\be\label{susyQS}
\{Q_a,S_b\}=2\epsilon_{ab}\mathcal{L}_1\,.
\ee
Here
\be\label{L1def}
\mathcal{L}_1=
\left(
\begin{array}{cc}
 \mathcal{P}^\alpha_1 & 0  \\
0 &  H_0\mathcal{P}_0
\end{array}
\right)
\ee
is the bosonic integral of motion composed from
the Lax-Novikov integrals of motion
of the subsystems $H^\alpha_1$ and $H_0$.
One  also can define  a bosonic operator
$\mathcal{L}_2=\sigma_3\mathcal{L}_1$.
It is another nontrivial bosonic integral
of the system,
\be\label{susyHL}
[\mathcal{H},\mathcal{L}_a]=0\,,
\ee
$a=1,2$. We have the following commutation relations
between bosonic, $\mathcal{L}_a$,  and fermionic,
$Q_a$ and $S_a$,
 integrals of motion
of the system $\mathcal{H}$\,:
\be\label{susyL1QS}
[\mathcal{L}_1,Q_a]=0\,,\qquad
[\mathcal{L}_1,S_a]=0\,,
\ee
\be\label{susyL2QS}
[\mathcal{L}_2,Q_a]=2i\mathcal{H}
S_a\,,\qquad
[\mathcal{L}_2,S_a]=-2i\mathcal{H}^2
Q_a\,.
\ee
The set of (anti)-commutation relations (\ref{susyHQS}),
(\ref{susyQQSS}), (\ref{susyQS}), (\ref{susyHL}),
(\ref{susyL1QS}) and (\ref{susyL2QS})
is the exotic nonlinear superalgebra
of the system (\ref{Hextend}).
The Lie superalgebra of  conventional $\mathcal{N}=2$ supersymmetry
is contained here as a sub-superalgebra
generated  by the Hamiltonian
$\mathcal{H}$ and the  first order supercharges $Q_a$.
The nonlinear extension of superalgebra
emerges here because the subsystem $H_0$
has its proper  integral of motion
$\mathcal{P}_0=-i\frac{d}{dx}$ that enters the structure of the
pair of additional supercharges $S_a$ which are
matrix differential operators of order two.
Like the Hamiltonian operator
$\mathcal{H}$, the bosonic integral $\mathcal{L}_1$ composed
from the Lax-Novikov integrals of motion
of the subsystems is a central element
of the superalgebra.
The second bosonic integral
$\mathcal{L}_2$ when commutes  with the pair
of supercharges $Q_a$ ($S_a$),
generates another pair of supercharges $S_a$ ($Q_a$).
The nonlinearity of the  superalgebra is related
to the appearance
of the central charge
$\mathcal{H}$ in the form of multipliers $\mathcal{H}$ and
$\mathcal{H}^2$ in the anticommutator of supercharges
$S_a$ and in commutation relations (\ref{susyL2QS})
of the supercharges with the bosonic integral $\mathcal{L}_2$.

The supersymmetry of the extended
 system  described by a Hamiltonian
$\mathcal{H}=\text{diag}\,(H^\alpha_n,H^\alpha_{n-1})$
composed from a  pair of
perfectly invisible neighbour $\mathcal{PT}$-symmetric
systems from the family
(\ref{HnPT})
has a similar structure.
Two supercharges $Q_a$ are matrix differential
operators of the first order which have the form
similar to (\ref{Qa}) with operators $D_1$ and $D^\#_1$
changed for the
operators $D_n$ and $D^\#_n$ defined in
(\ref{Dn}). The supercharges $S_a$
are matrix differential operators of order $2n$.
They have a structure similar to that of the
supercharges  (\ref{Sa}) with
the operators $D_1 \mathcal{P}_0$ and
$\mathcal{P}_0 D^\#_1$ changed for
$D_n  \mathcal{P}_{n-1}^\alpha$
and $ \mathcal{P}_{n-1}^\alpha D^\#_n$,
respectively,
where $\mathcal{P}_{n-1}^\alpha$ is the Lax-Novikov
integral of the subsystem $H^\alpha_{n-1}$ given by
Eq. (\ref{Pn}).  The analog
of the bosonic integral (\ref{L1def})
has in this case the form
$\mathcal{L}_1=\text{diag}\,(\mathcal{P}_{n}^\alpha,
H^\alpha_{n-1}\mathcal{P}_{n-1}^\alpha)$, and, again,
another nontrivial bosonic integral is
$\mathcal{L}_2=\sigma_3\mathcal{L}_1$.
The anti-commutator of  supercharges $Q_a$
with $Q_b$  has  the same form as in (\ref{susyQQSS}),
the anticommutator of $Q_a$ with $S_b$ has the form
(\ref{susyQS}).
The anti-commutator between supercharges
$S_a$ and $S_b$
changes here for
$\{S_a,S_b\}=2\delta_{ab}(\mathcal{H})^{2n}$,
which is a corresponding generalization
of the second relation from (\ref{susyQQSS}).
Here $\mathcal{L}_1$ is again a central element
of superalgebra generated via the anti-commutator
between  supercharges $Q_a$ and $S_b$.
The commutator of $\mathcal{L}_2$ with $Q_a$
has exactly the same form as in (\ref{susyL2QS}),
while the commutator $\mathcal{L}_2$ with $S_a$
is changed  for
$[\mathcal{L}_2,S_a]=-2i\mathcal{H}^{2n}Q_a$,
that is a natural generalization of the second relation from
(\ref{susyL2QS}).
Note that the matrix differential operator $K$ from Eq. (\ref{operK}) that controls 
the Toda kink fluctuations, after 
$\mathcal{PT}$-regularization $x\rightarrow x+i\alpha$ 
corresponds exactly to the case $n=2$
of  the supersymmetric Hamiltonian
$\mathcal{H}=\text{diag}\,(H^\alpha_n,H^\alpha_{n-1})$
with permuted  subsystems.
So, the system of $\mathcal{PT}$-regularized kinks of the 
$SU(3)$  conformal Toda system is described 
by the exotic  $\mathcal{N}=4$ nonlinear supersymmetry.

In the same way one can identify the superalgebra
of the extended systems composed from the pairs
of the systems $H^{(1)}_{\alpha,\gamma}$ and $H^\alpha_1$,
and of the systems $H^{(2)}_{\alpha,\gamma}$ and $H^\alpha_2$,
where $H^{(1)}_{\alpha,\gamma}$ and $H^{(2)}_{\alpha,\gamma}$
are the perfectly invisible
$\mathcal{PT}$-symmeric systems that we described at the end
of Section \ref{Section2}.

We also can consider  the exotic supersymmetry
of an  extended system composed from any two
perfectly invisible $\mathcal{PT}$-symmetric systems
by taking into account that each such a system is characterized
by a nontrivial Lax-Novikov integral being
a Darboux-dressed momentum operator of the free particle. As a consequence,
the corresponding  two subsystems always can be intertwined by  two distinct
differential operators, see for 
such a structure appearing  in $n$-soliton \cite{ArMatPly1}
and periodic finite-gap \cite{CJNP} systems.

\vskip0.1cm

Like in the case of reflectionless  
systems \cite{ArMatPly1,PlyuNie,AraPly},
sometimes there could appear the operators which intertwine
the two perfectly invisible $\mathcal{PT}$-symmeric systems
directly but not via the
chains of the intertwining operators ascending to the free particle system.
Then the structure of the superymmetry algebra will
transmute and include
some central element  composed from  the complex shift parameters.

As an example where such a situation is realized,
let us consider the extended system
\be\label{Ha1a2}
\mathcal{H}=
\left(
\begin{array}{cc}
 H^{\alpha_2}_1 & 0   \\
0  &  H^{\alpha_1}_1
\end{array}
\right)
\ee
composed from two systems $H^{\alpha_1}_1$ and $H^{\alpha_2}_1$
of the form (\ref{H1PT})
characterized by two different nonzero  shift parameters $\alpha_1$
and $\alpha_2$. For the sake of definiteness,
we assume that $\alpha_1>\alpha_2$.
The two subsystems $H^{\alpha_1}_1$ and $H^{\alpha_2}_1$
can be intertwined by the second
order differential operators $D_{\alpha_1}D_{\alpha_2}^\#$ and
$D_{\alpha_2}D_{\alpha_1}^\#$, where
$D_{\alpha_j}$ and $D_{\alpha_j}^\#$, $j=1,2$,
are the first order differential operators
of the form (\ref{D1}),
$D_{\alpha_j}=\frac{d}{dx}-(x+i\alpha_j)^{-1}$,
$D_{\alpha_j}^\#=-\frac{d}{dx}-(x+i\alpha_j)^{-1}$,
each of which intertwines a subsystem $H^{\alpha_j}_1$
with a free particle $H_0$, see Eq. (\ref{HaH0inter}).
So, we have here the intertwining relations
\be
(D_{\alpha_1}D_{\alpha_2}^\#)H^{\alpha_2}_1=
H^{\alpha_1}_1(D_{\alpha_1}D_{\alpha_2}^\#)\,,\qquad
(D_{\alpha_2}D_{\alpha_1}^\#)H^{\alpha_1}_1=
H^{\alpha_2}_1(D_{\alpha_2}D_{\alpha_1}^\#)\,.
\ee
These second order differential operators
intertwine the subsystems $H^{\alpha_1}_1$
and $H^{\alpha_2}_1$ effectively via
a `virtual' free particle  system $H_0$\,:
$(D_{\alpha_1}D_{\alpha_2}^\#)H^{\alpha_2}_1=
D_{\alpha_1}(D_{\alpha_2}^\#H^{\alpha_2}_1)=(D_{\alpha_1}
H_0)D_{\alpha_2}^\#=H^{\alpha_1}_1(D_{\alpha_1}D_{\alpha_2}^\#)$.
Besides them,  we also have here the intertwiners
\be\label{DW12}
D=\frac{d}{dx}+\mathcal{W}\,,\qquad
D^\#=-\frac{d}{dx}+\mathcal{W}\,,
\ee
where
\be\label{Wdef}
\mathcal{W}=\frac{1}{\xi_1}-\frac{1}{\xi_2}-
\frac{1}{\xi_1-\xi_2},
\ee
$\xi_j=x+i\alpha_j$, and
$-(\xi_1-\xi_2)^{-1}=i(\alpha_1-\alpha_2)^{-1}$.
The first order differential operators (\ref{DW12})
 intertwine the subsystems directly,
\be
DH^{\alpha_1}_1=H^{\alpha_2}_1D\,,\qquad
D^\#H^{\alpha_2}_1=H^{\alpha_1}_1D^\#\,,
\ee
and also satisfy the relations
\be
D^\# D=H^{\alpha_1}_1-\Delta^2\,,\qquad
DD^\#=H^{\alpha_2}_1-\Delta^2\,,
\ee
where we have introduced the notation
\be
\Delta=\frac{1}{\alpha_1-\alpha_2}\,.
\ee
When $\alpha_1\rightarrow \infty$, the Hamiltonian
operator
$H^{\alpha_1}_1$ transforms into $H_0$,
for function (\ref{Wdef}) we have 
$\mathcal{W}\rightarrow -\xi_2^{-1}$, and
$D$ transforms into the operator $D_1$
given by Eq. (\ref{D1}) with $\xi$ changed for $\xi_2$,
that intertwines  the $H_0$ with $H^{\alpha_2}_1$.

To establish the explicit form of the
exotic superalgebraic structure of the system
(\ref{Ha1a2}), the following relations that involve
the  operators (\ref{DW12}), and $D_{\alpha_j}$ and
$D^\#_{\alpha_j}$, $j=1,2$,  are useful\,:
\be
DD_{\alpha_1}=D_{\alpha_2}D_0\,,\qquad
D^\#_{\alpha_2}D=D_0D^\#_{\alpha_1}\,,
\ee
\be
D^\#D_{\alpha_2}=D_{\alpha_1}D^\#_0\,,\qquad
D^\#_{\alpha_1}D^\#=D^\#_0D^\#_{\alpha_2}\,,
\ee
where we denoted
\be
D_0\equiv \frac{d}{dx}+i\Delta\,,\qquad
D^\#_0\equiv -\frac{d}{dx}+i\Delta\,.
\ee
Note that the first order  operators (\ref{DW12})
and the second order operators $D_{\alpha_1}D_{\alpha_2}^\#$ and
$D_{\alpha_2}D_{\alpha_1}^\#$ intertwine also the
Lax-Novikov integrals of the subsystems,
\be
D\mathcal{P}^{\alpha_1}=\mathcal{P}^{\alpha_2}D\,,\qquad
D^\#\mathcal{P}^{\alpha_2}=\mathcal{P}^{\alpha_1}D^\#\,,
\ee
\be
(D_{\alpha_2}D_{\alpha_1}^\#)\mathcal{P}^{\alpha_1}=
\mathcal{P}^{\alpha_2}(D_{\alpha_2}D_{\alpha_1}^\#)\,,\qquad
(D_{\alpha_1}D_{\alpha_2}^\#)\mathcal{P}^{\alpha_2}=
\mathcal{P}^{\alpha_1}(D_{\alpha_1}D_{\alpha_2}^\#)\,.
\ee

{}From the described intertwining relations 
we find that the system (\ref{Ha1a2})  is characterized by
the supercharges
\be
Q_1= \left(
\begin{array}{cc}
0 & D   \\
D^\#  &  0
\end{array}
\right),\qquad
Q_2=i\sigma_3Q_1\,,
\ee
which are the matrix first order differential operators,
and by the supercharges
 \be
S_1= \left(
\begin{array}{cc}
0 & D_{\alpha_2}D^\#_{\alpha_1}   \\
D_{\alpha_1} D^\#_{\alpha_2}&  0
\end{array}
\right),\qquad
S_2=i\sigma_3S_1\,,
\ee
which are matrix differential operators of the second order.
Besides, the extended system  (\ref{Ha1a2})
has two nontrivial bosonic integrals of motion
 \be
\mathcal{L}_1= \left(
\begin{array}{cc}
\mathcal{P}^{\alpha_2} & 0   \\
0&  \mathcal{P}^{\alpha_1}
\end{array}
\right),\qquad
\mathcal{L}_2=\sigma_3\mathcal{L}_1\,.
\ee
All these integrals of motion generate the following
exotic nonlinear superalgebra of the system
(\ref{Ha1a2})\,:
\be
[\mathcal{H},Q_a]=0\,,\qquad
[\mathcal{H},S_a]=0\,,\qquad
[\mathcal{H},\mathcal{L}_a]=0\,,
\ee
\be\label{QQSSbr}
\{Q_a,Q_b\}=2\delta_{ab}(\mathcal{H}-\Delta^2)\,,\qquad
\{S_a,S_b\}=2\delta_{ab}\mathcal{H}^2\,,
\ee
\be
\{Q_a,S_b\}=2\left(\epsilon_{ab}\mathcal{L}_1+i\delta_{ab}\Delta\mathcal{H}
\right)\,,
\ee
\be
[\mathcal{L}_a,\mathcal{L}_b]=0\,,\qquad
[\mathcal{L}_1,Q_a]=0\,,\qquad
[\mathcal{L}_1,S_a]=0\,,
\ee
\be
[\mathcal{L}_2,Q_a]=2i(\mathcal{H}-\Delta^2)S_a+
2\Delta\cdot \mathcal{H}Q_a\,,
\qquad 
[\mathcal{L}_2,S_a]=-2i
\mathcal{H}^2Q_a-2
\Delta\cdot \mathcal{H}S_a\,.
\ee

In the limit $\alpha_1\rightarrow \infty$,
the system
(\ref{Ha1a2}) transforms into the system (\ref{Hextend}),
and all the integrals of (\ref{Ha1a2}) are transformed
into  the corresponding  integrals of motion of the system
(\ref{Hextend}) with parameter $\alpha$ changed for $\alpha_2$.
The exotic superalgebra of the system (\ref{Ha1a2})
in this limit  takes the form of the exotic superalbegra
of the system (\ref{Hextend}).

\vskip0.1cm
In accordance with distinct  exotic nonlinear 
superalgebraic structures 
of the supersymmetric systems 
(\ref{Hextend}) and (\ref{Ha1a2}), their spectral properties 
also essentially differ.
Consider  first the system 
(\ref{Hextend}).  Its integrals $\mathcal{H}$,
$\mathcal{L}_a$, $Q_2$ and $S_1$ are $\mathcal{PT}$-even
operators, while $Q_1$ and $S_2$ are  $\mathcal{PT}$-odd\,:
\be\label{PTparity}
{PT}\mathcal{E}=\mathcal{E}{PT}\,,\quad
\mathcal{E}=\mathcal{H}\,,\mathcal{L}_a\,,
Q_2\,,S_1\,;\qquad
{PT}\mathcal{O}=-\mathcal{O}{PT}\,,\quad
\mathcal{O}=
Q_1\,,S_2\,.
\ee
For fermionic integrals of this system the 
following operator  identities are valid
coherently with (\ref{susyQS})\,:
 $Q_1S_1=-S_1Q_1=Q_2S_2=-S_2Q_2=i\mathcal{L}_2$,
$Q_1S_2=S_2Q_1=-Q_2S_1=-S_1Q_2=\mathcal{L}_1$.
The first order
supercharges $Q_1$  and $Q_2$ commute, respectively,  with 
the second order  supercharges $S_2$  and $S_1$,
and the corresponding pairs of the fermionic integrals 
 can be diagonalized simultaneously.
The functions 
\be\label{Psik}
\Psi_k^\pm(x)=
\left(
\begin{array}{cc}
  D_1e^{ikx} \\
   \mp ike^{ikx} \\
\end{array}
\right),\qquad
-\infty<k<\infty\,,
\ee
are the common eigenstates of $\mathcal{H}$, $\mathcal{L}_1$,
$Q_2$ and $S_1$\,:
\be
\mathcal{H}\Psi_k^\pm=k^2\Psi_k^\pm\,,\quad
\mathcal{L}_1\Psi_k^\pm=k^3\Psi_k^\pm\,,\quad
Q_2\Psi_k^\pm=\pm k\Psi_k^\pm\,,\quad
S_1\Psi_k^\pm=\mp k^2 \Psi_k^\pm\,.
\ee
The integral $\mathcal{L}_1$ 
distinguishes the eigenstates with positive and negative values of $k$,
while both supercharges $Q_2$ and  $S_1$
separate the eigenstates with different values $+$ and $-$ of 
the upper  index.
On the other hand, 
$Q_1\Psi_k^\pm=\pm k(i\sigma_3\Psi_k^\pm)$,
$S_2\Psi_k^\pm=\pm k^2(i\sigma_3\Psi_k^\pm)$,
$\mathcal{L}_2\Psi_k^\pm=k^3(\sigma_3\Psi_k^\pm)$\,.
The  state with $k=0$,  $\Psi_0(x)=
( D_1 1, 0)^t =(-\xi^{-1},0)^t$, where $t$ means a transposition,
is a unique eigenstate of zero energy that is annihilated by all the integrals 
of motion.  The $\mathcal{PT}$-symmetric system (\ref{Hextend})  corresponds therefore to the 
case of unbroken exotic nonlinear 
supersymmetry.
The  commuting $\mathcal{PT}$-odd supercharges 
$Q_1$ and $S_2$  have the common  eigenfunctions
$\widetilde{\Psi}^\pm_k(x)=(D_1e^{ikx},\pm k e^{ikx})^t$
with real eigenvalues,
$Q_1\widetilde{\Psi}^\pm_k=\pm k \widetilde{\Psi}^\pm_k$,
$S_2\widetilde{\Psi}^\pm_k=\pm k^2 \widetilde{\Psi}^\pm_k$.
Analogously to the states (\ref{Psik}), 
they are eigenfunctions of $\mathcal{H}$ and $\mathcal{L}_1$
with eigenvalues $k^2$ and $k^3$, respectively, 
and satisfy 
the relations $Q_2\widetilde{\Psi}^\pm_k=\pm k(i\sigma_3\widetilde{\Psi}^\pm_k)$,
$S_1\widetilde{\Psi}^\pm_k=\mp k^2(i\sigma_3\widetilde{\Psi}^\pm_k)$,
$\mathcal{L}_2\widetilde{\Psi}^\pm_k= k^3(\sigma_3\widetilde{\Psi}^\pm_k)$.

\vskip0.1cm
Consider now the system  (\ref{Ha1a2}).
Its integrals 
satisfy the same parity properties (\ref{PTparity}) 
with respect to permutations with the operator $PT$.
Here we have the identities 
$Q_1S_1=Q_2S_2=i(\sigma_3 \mathcal{L}_1+\Delta\mathcal{H})$,
$S_1Q_1=S_2Q_2=i(-\sigma_3 \mathcal{L}_1+\Delta\mathcal{H})$,
$Q_1S_2=-Q_2S_1= \mathcal{L}_1+\Delta\sigma_3\mathcal{H}$,
$S_2Q_1=-S_1Q_2= \mathcal{L}_1-\Delta\sigma_3\mathcal{H}$.
Unlike (\ref{Hextend}),  in the system  (\ref{Ha1a2}) 
there are no mutually commuting pairs
of the first and second order supercharges,
and  the complete sets of mutually commuting integrals are formed 
by the sets of the bosonic 
operators $\mathcal{H}$ and  $\mathcal{L}_1$ supplemented 
by one of the fermionic first order supercharges $Q_a$,
or by one of the second order supercharges $S_a$.

The common eigenfunctions of the set 
of the $\mathcal{PT}$-even operators ($\mathcal{H}$, $\mathcal{L}_1$, $S_1$)
are  
\be\label{PsiDk2}
\Psi^\pm_{k^2}=
\left(
\begin{array}{cc}
 D_{\alpha_2}e^{ikx} \\
 \pm D_{\alpha_1}e^{ikx} \\
\end{array}
\right),\qquad
-\infty<k<\infty\,,
\ee
for which $S_1\Psi^\pm_{k^2}=\pm k^2\Psi^\pm_{k^2}$,
$\mathcal{H}\Psi^\pm_{k^2}=k^2\Psi^\pm_{k^2}$,
$\mathcal{L}_1\Psi^\pm_{k^2}=k^3\Psi^\pm_{k^2}$.
They also satisfy the relations 
$Q_1\Psi^\pm_{k^2}=\pm i(k\sigma_3+\Delta)\Psi^\pm_{k^2}$,
$Q_2\Psi^\pm_{k^2}=\mp (k+\sigma_3\Delta)\Psi^\pm_{k^2}$,
$S_2\Psi^\pm_{k^2}=\pm k^2(i\sigma_3\Psi^\pm_{k^2})$,
$\mathcal{L}_2\Psi^\pm_{k^2}=k^3(\sigma_3\Psi^\pm_{k^2})$.
The doublet of states 
$\Psi^\pm_{0}=(D_{\alpha_2}1,\pm D_{\alpha_1}1)^t=
(-\xi^{-1}_2,\mp\xi^{-1}_1)^t$ corresponding to $k=0$ are  the eigenfunctions 
with the lowest, zero  value of energy,
 $\mathcal{H}\Psi^\pm_{0}=0$,  which also 
are annihilated by both bosonic integrals $\mathcal{L}_a$
and both second order supercharges $S_a$.
However, they are not annihilated by the first order supercharges,
being eigenfuctions of the $Q_1$ with imaginary eigenvalues\,:
$Q_1\Psi^\pm_{0}=\pm i\Delta\Psi^\pm_0$.
Their linear combinations  $\Psi^+_0 \pm i\Psi^-_0$ are eigenfunctions 
of the $\mathcal{PT}$-even first order supercharge $Q_2$ 
with the same pure imaginary eigenvalues $\pm i\Delta$.
Unlike (\ref{Hextend}),
the $\mathcal{PT}$-symmetric system  (\ref{Ha1a2}) realizes the case 
of  spontaneously partially broken exotic nonlinear supersymmetry.
Analogously, the states $\widetilde{\Psi}^\pm_{k^2}=
(D_{\alpha_2}e^{ikx},\pm i D_{\alpha_1}e^{ikx})^t$
are common eigenstates of $\mathcal{H}$, $\mathcal{L}_1$
and $S_2$, 
$S_2\widetilde{\Psi}^\pm_{k^2}=\pm k^2\widetilde{\Psi}^\pm_{k^2}$,
$\mathcal{H}\widetilde{\Psi}^\pm_{k^2}=k^2\widetilde{\Psi}^\pm_{k^2}$,
$\mathcal{L}_1\widetilde{\Psi}^\pm_{k^2}=k^3\widetilde{\Psi}^\pm_{k^2}$,
for which we also have the relations 
$\mathcal{L}_2\widetilde{\Psi}^\pm_{k^2}=k^3(\sigma_3\widetilde{\Psi}^\pm_{k^2})$,
$S_1\widetilde{\Psi}^\pm_{k^2}=-\mp k^2(i\sigma_3\widetilde{\Psi}^\pm_{k^2})$,
$Q_1\widetilde{\Psi}^\pm_{k^2}=
\mp(k+\Delta\sigma_3)\widetilde{\Psi}^\pm_{k^2}$,
$Q_2\widetilde{\Psi}^\pm_{k^2}=\mp i(k\sigma_3+\Delta)\widetilde{\Psi}^\pm_{k^2}$.
The states $\widetilde{\Psi}^\pm_0$ are annihilated 
by $S_a$, $\mathcal{H}$ and $\mathcal{L}_a$,
but they are eigenfunctions of $Q_2$ of nonzero,  pure imaginary eigenvalues
$\mp i\Delta$, and their linear combinations 
$\widetilde{\Psi}^+_0\pm i\widetilde{\Psi}^-_0$ 
are eigenstates of $Q_1$ of eigenvalues $\pm i\Delta$.

According to the first relation from (\ref{QQSSbr}), the 
eigenvalues of the first order supercharges $Q_a$
have to be pure imaginary for $0\leq k^2<\Delta^2$, equal to zero for $k^2=\Delta^2$, and 
to be real for energy eigenvalues $k^2>\Delta^2$.
In accordance with this, the eigenstates  of $Q_1$ with eigenvalues $\lambda=\pm i\sqrt{\Delta^2-k^2}$
are $\Psi^\pm_k(x)=(\sqrt{\Delta+k}D_{\alpha_2}e^{ikx},\,
\mp \sqrt{\Delta-k}D_{\alpha_1}e^{ikx})^t$  for $0\leq k^2<\Delta^2$. Particularly, the states 
$\Psi^\pm_0=(D_{\alpha_2}1,\,\pm D_{\alpha_2}1)^t$ are two eigenstates of 
eigenvalues $\pm i\Delta$ of zero energy that already have  been 
described above. For $k^2>\Delta^2$, the eigenstates of $Q_1$ of 
real eigenvalues $\lambda=\pm\sqrt{k^2-\Delta^2}$ have the form 
$\Psi^\pm_k(x)=(\sqrt{\vert \Delta+k\vert }D_{\alpha_2}e^{ikx},\,
\mp i\varepsilon_k\sqrt{\vert k-\Delta\vert }D_{\alpha_1}e^{ikx})^t$,
where $\varepsilon_k$ denotes the sign of $k$.
These states $\Psi^\pm_k(x)$ also are eigenstates of $\mathcal{H}$ and 
$\mathcal{L}_1$ of eigenvalues $k^2$ and $k^3$, respectively.
The eigenstates of the $\mathcal{PT}$-even supercharge $Q_2$
can be obtained from the eigenstates of $Q_1$.
Denoting by  $\Psi_\lambda$ an eigenstate of $Q_1$ 
of eigenvalue $\lambda$, $Q_1\Psi_\lambda=\lambda\Psi_\lambda$,
whose explicit structure for different values of $\lambda$ we have just 
described, the eigenstate of $Q_2$ of eigenvalue $-\lambda$ is given by 
$\widetilde{\Psi}_{-\lambda}=(\Pi_++i\Pi_-)\Psi_\lambda$,
$Q_2\widetilde{\Psi}_{-\lambda}=-\lambda\widetilde{\Psi}_{-\lambda}$.
Here $\Pi_\pm=\frac{1}{2}(1\pm\sigma_3)$ are the projectors.

In the limit $\alpha_1\rightarrow \infty$, we have $\Delta\rightarrow 0$, 
and the described phase of the partially 
broken exotic nonlinear supersymmetry of the system 
(\ref{Ha1a2}) transmutes  into the phase of unbroken 
exotic nonlinear supersymmetry of the system (\ref{Hextend}).

\section{Summary, discussion and outllok}\label{Section5}

We investigated  a special class of  $\mathcal{PT}$-symmetric
quantum mechanical systems. Being generated via 
$\mathcal{PT}$-regularized 
Darboux-Crum  transformations of
the free particle system $H_0$, each of them has the same spectrum as $H_0$
but with a bound state $\psi_{0,0}(x)=1$ of zero energy of $H_0$
transformed into a bounded (quadratically integrable on $\R^1$) eigenstate of the same energy
of the generated system. Each such a system satisfies a stationary equation 
of the KdV hierarchy and is characterized by a Lax-Novikov integral
being  a higher order differential operator.
It is a Darboux-dressed free particle momentum integral 
which distinguishes the left- and right-moving 
scattering states inside the spectral doublets 
and detects the bounded state of zero energy
by annihilating it. Other states from the kernel of this integral 
are the Jordan states of the corresponding system.
The peculiarity of such systems is that 
in them not only the reflection coefficient is zero, and so, 
their transmission coefficient is a pure phase, but that 
this phase is equal to one like for the free particle.
We identify  this class of the quantum systems 
as perfectly invisible 
$\mathcal{PT}$-symmetric zero-gap systems.
Their family  includes as particular cases 
the $\mathcal{PT}$-regularized
two-particle Calogero systems 
(conformal quantum mechanics models
of  de Alfaro-Fubini-Furlan)
 with coupling constant values 
 $g=n(n+1)$.
 The generation of the 
 $\mathcal{PT}$-symmetric zero-gap systems
 more complicated than the simplest case of 
 the $\mathcal{PT}$-regularized Calogero
 model with $n=1$ includes Jordan states
 of the free particle taken as the seed states
 in the Darboux-Crum construction.
 
In some  systems (different from the 
 Calogero case) we observed an interesting 
 phenomenon of the mutual transmutation  of 
 physical and non-physical eigenstates of zero energy
 in the limit when one of the imaginary parameters 
 is sent to infinity. The rational potentials of
perfectly invisible 
$\mathcal{PT}$-symmetric zero-gap systems, 
being solutions of stationary equations
of the KdV hierarchy,  can be promoted to the solutions 
of  time-dependent  equations from 
the same hierarchy. 
The interesting 
peculiarity of time-dependent solutions 
we constructed is that under appropriate 
choice of the parameters  they reveal 
a behaviour being typical for  extreme 
waves.
It is worth noting here that if
 we separate a 
$\mathcal{PT}$-symmetric function $u(x,t)$
in real and imaginary parts, $u(x,t)=v(x,t)+iw(x,t)$,
in the case of the KdV equation 
we obtain an equivalent system
of two coupled equations 
\be\label{vw}
v_t-3(v^2-w^2)_x+v_{xxx}=0\,,\qquad
w_t-6(vw)_x+w_{xxx}=0
\ee
for two real fields $v(x,t)$ and $w(x,t)$ instead of 
one equation for the 
complex-valued field $u(x,t)$.
At $w=0$, this system reduces
to the KdV equation, while for $v=0$ it corresponds 
to the case of the linearized KdV equation.
We refer here  to \cite{CaFriBag,CenFring}
where different aspects of 
the system (\ref{vw}) 
 were discussed in the context of the complex KdV 
 equation.

We showed that the two simplest Hamiltonians 
from the Calogero subfamily with $n=1$ and $n=2$ 
governing the dynamics
in conformal invariant $\mathcal{PT}$-symmetric 
quantum mechanical systems also 
determine the fluctuation spectra around the singular
kinks arising as traveling waves in the field-theoretical 
Liouville and $SU(3)$ 
conformal Toda systems.
By pushing  the centers of the  kinks to the imaginary axis 
we regularize them and the operators governing 
the fluctuations around them take the form 
of the Hamiltonians of the corresponding 
$\mathcal{PT}$-regularized Calogero systems.
 This picture can be  
related to  the contraction of 
the infinite-dimensional conformal group in 
the two-dimensional Minkowskian setting to a
finite subgroup in conformal quantum mechanics.

The peculiar properties of the 
investigated class of the systems are reflected 
in the quantum mechanical supersymmetry associated with them.
In the extended systems composed from the 
pairs of such systems related by a  first order
Darboux 
transformation,  the  conventional  $\mathcal{N}=2$ supersymmetry
is extended to exotic $\mathcal{N}=4$  nonlinear supersymmetry.
The latter includes an additional pair
of supercharges to be matrix 
differential operators of the even order constructed 
from the higher order  differential operators which intertwine
the Hamiltonians of the subsystems via
a `virtual' free particle system. As a result, the first 
and higher order supercharges  generate the 
Lax-Novikov integrals of the subsystems which
compose one of the additional bosonic integrals 
of the extended system  being a central charge $\mathcal{L}_1$
of the superalgebra alongside with the Hamiltonian
operator $\mathcal{H}$.
Another  additional bosonic integral corresponds 
to the operator $\mathcal{L}_2=\sigma_3\mathcal{L}_1$
that generates a rotation between the first  
and higher order supercharges with coefficients
to be powers  of the Hamiltonian.
The appearance of the Hamiltonian in the structure 
coefficients of the superalgebra corresponds to a 
nonlinear nature of the exotic 
supersymmetry of  such systems.
This described family of the extended  systems realizes 
the case of the unbroken exotic supersymmetry in which
there is one singlet state of zero eigenvalue of $\mathcal{H}$
which is annihilated by all the four  supercharges and by both 
additional bosonic integrals.
The system of the $\mathcal{PT}$-regularized kinks of the 
$SU(3)$  conformal Toda system can be described 
by such unbroken exotic  $\mathcal{N}=4$ nonlinear supersymmetry.
There is also another family of the extended 
two-component systems 
corresponding to the case of the partially
broken exotic supersymmetry
in which a doublet of zero energy states
is annihilated by the bosonic Lax integrals 
and higher-order supercharges but is
not annihilated by the first order supercharges.
We considered a simplest  example of such a 
supersymmetric system in which the first order
supercharges are composed from the first order 
operators which intertwine the completely isospectral
subsystems directly, 
while the intertwining operators in the structure of 
the pair of the second order supercharges relate 
the subsystems via a virtual free particle system.
In a certain limit such a system with partially broken 
exotic supersymmetry transmutes into 
the perfectly invisible extended system
described by unbroken supersymmetry.

Having in mind the analogy with the case of 
reflectionless  quantum systems related to the   conventional
multi-soliton 
solutions of the KdV \cite{AraPly}, 
one can expect that there exists
 an infinite chain  of the 
the  pairs of the  perfectly invisible 
$\mathcal{PT}$-symmetric zero-gap systems
in which one of the  neighbour  pairs is
described by unbroken exotic supersymmetry
while another neighbour pair is characterized by partially broken
phase of the exotic nonlinear supersymmetry
and the neighbour pairs with different phases 
of supersymmetry can be transmuted via the process
of soliton scattering. It would be interesting 
to construct  the indicated infinite chain 
of the  pairs of the  perfectly invisible 
$\mathcal{PT}$-symmetric zero-gap
Schr\"odinger systems. One of the first order supercharges
 in such chains of the systems can then be considered as 
 a Hamiltonian of a (1+1)-dimensional  
 Dirac system with a perfectly invisible 
 $\mathcal{PT}$-symmetric  scalar potential
 that descirbes a  fermion in a 
 multi-kink-antikink background \cite{AraMatPly2}.

The $\mathcal{PT}$-regularized perfectly invisible 
two-particle Calogero systems we considered  
possess conformal symmetry like their 
 conventional counterparts with real inverse square 
potentials \cite{deAFF}. The interesting question is what
happens with conformal symmetry
in the perfectly invisible $\mathcal{PT}$-symmetric 
zero-gap systems of different form,
like in the systems described by the 
$\mathcal{PT}$-symmetric potentials (\ref{U+}) and 
(\ref{V+10}). Another, related question is
what happens with the exotic nonlinear 
supersymmetry we considered 
under extension of it by taking into account 
the conformal symmetry.
A priori it is not clear whether in such a case 
we obtain some finite nonlineaer superalgebraic structure or
some infinite-dimensional superalgebraic structure will be 
generated,
cf. \cite{FedIvaLec,IvanKriLec,LeiPly}. 
We are going to consider  these questions 
elsewhere \cite{MatPly+}. 

The interesting question 
 also
  is whether
 perfectly invisible $\mathcal{PT}$-symmetric 
zero-gap systems to  be more complicated
than the $\mathcal{PT}$-regularized 
Calogero systems with $n=1$ and $n=2$
can be related  to the field-theoretical  conformal models.

In conclusion we note that in \cite{BorMat}
there were considered self-similar solutions of the KdV equation
with initial conditions 
of the form
$N(N+1)/x^2$, which are rational singular solutions on 
the real line. Following \cite{MatPositons},
one could consider the solutions to the 
Schr\"odinger problem with such a potential 
on half-line $x>0$ (see Appendix)  and then formally
extend  the same solutions to another side  $x<0$ 
of the pole as it is done in \cite{MatPositons} for 
the so called
positon solutions. 
With such a formal treatment of solutions
one could arrive at the conclusion
that the indicated 
 rational potentials are reflectionless on all the real line,
and moreover, that they would be characterized 
by the transmission amplitude equal to one.
Treated in this way singular potentials  
are called in  \cite{MatPositons} as 
``super-transparent'' or ``super-reflectionless".
However, physically such a treatment of 
singular potentials on all the real line is 
completely formal since the second order pole singularity at $x=0$,
unlike  the case of delta function potential,   is not penetrable,
and  the physically admissible states  in the 
regions $x<0$ and $x>0$ do not mix dynamically.  
The  impermissibility of such 
treatment of the quantum problem also is reflected  by 
the mentioned in 
Section \ref{SecIntro} observation that the 
Lax-Novikov operator commuting with Hamiltonian 
in this case is not observable 
since it  transforms physical states into
non-physical ones \cite{COPl}.
It is the $\mathcal{PT}$-symmetric 
regularization considered here that radically changes 
the properties of such systems
and their interpretation.
It would be interesting to apply 
the $\mathcal{PT}$-symmetric 
regularization 
to the generalized Darboux transformations
used in  \cite{MatPositons}
for the construction of time-dependent
positon and soliton-positon solutions of the KdV equation.

\vskip0.2cm
 
\noindent {\large{\bf Acknowledgements} } 
\vskip0.3cm

We acknowledge support from research projects
FONDECYT 1130017 and
Convenio Marco Universidades del Estado (Project USA1555), Chile, 
and MINECO (Project MTM2014-57129-C2-1-P), Spain.
JMG also acknowledges the Junta de Castilla 
y Le\'on for financial support under grant VA057U16.
MSP is grateful for the warm hospitality at  Salamanca and Valladolid Universities
where a part of this work was  done.

\section{Appendix}\label{Section6}

Here we briefly discuss  a general picture of the 
quantum scattering problem on the half-line.

Consider a free particle of mass $m=\frac{1}{2}$ on the half-line
$(0,\infty)$ 
given  by the Hamiltonian operator
\be\label{HhalfU}
H=-\frac{d^2}{dx^2}+U(x)
\ee
with $U(x)=0$ for $x>0$ and $U(x)=+\infty$ for $x\leq 0$.
Its bounded eigenstates  of energy $E=k^2>0$ with  $k>0$ are 
described by wave functions 
$\psi_k(x)=C \sin kx$ for $x>0$ and $\psi_k(x)=0$ for $x\leq 0$.
They  represent  
a linear combination of two plane waves,
$\psi_k(x)=e^{-ikx}-e^{ikx}$, in which   the first term
can be interpreted as an incident from $x=+\infty$
plane wave, while the second
term can be considered as a reflected
wave. 

Let us consider now the case of a quantum 
particle on the half-line $(0,\infty)$ subject to a 
nontrivial potential $U(x)$ that
disappears sufficiently rapidly at $x= +\infty$, and 
$U(x)\rightarrow +\infty$ when $x\rightarrow 0$
in such a way  that 
the particle cannot penetrate  
to the negative half-line\,:  $\psi(0)=0$ for $x\leq 0$.
Then for  eigenfunction  corresponding to
eigenvalue $E=k^2>0$ the probability current 
$j(x)\propto \psi_k^*(x)(\psi_k(x))'-(\psi_k^*(x))' \psi_k(x)$
disappears when $x\rightarrow 0^+$\,: $j(0^+)=0$.
For $x\rightarrow +\infty$,  
we fix  the asymptotic form of the solution of the equation 
$H\psi_k(x)=k^2\psi_k(x)$ 
in  a form
\be\label{rinfty}
\psi_k(x)=e^{-ikx}+re^{ikx},
\ee
 where $r$ is a complex constant that
can be interpreted as a reflection amplitude.
The continuity equation $\frac{\partial}{\partial t}\rho+\frac{\partial}{\partial x}j=0$
for stationary solutions of the corresponding time-dependent 
Schr\"odinger equation 
reduces to the condition
$\frac{\partial}{\partial x}j=0$. Since $j(0^+)=0$, 
the probability current $j(x)=0$ for any $x\in(0,\infty)$.
 For asymptotic form of the solution 
 (\ref{rinfty}) this gives $rr^*=1$, i.e. the 
 reflection amplitude $r$ is a pure phase.
 In the case of the free paticle on the half-line 
 we have $r=-1$.
In the case of the potential $U(x)=n(n+1)/x^2$,
the solutions are given by the 
Darboux-Crum mapping  of the 
free particle solutions\,:
$\psi^{(n)}_k(x)=\mathcal{D}_n\sin kx$,
where $\mathcal{D}_n=D_n\ldots D_1$,
$D_j=\frac{d}{dx}-\frac{j}{x}$, $j=1,\ldots,n$.
These eigenfunctions $\psi^{(n)}_k(x)$ and their derivatives up to (including) 
the order $n$ annul at $x=0$. 
 For $x\rightarrow+\infty$, the asymptotic form of eigenfunctions is
$\psi^{(n)}_k(x)=-(-ik)^n(e^{-ikx}+(-1)^{n+1} e^{ikx})$.
{}From here  we find that 
the system with potential
$U(x)=n(n+1)/x^2$ is charcterized by
reflection amplitude $r=(-1)^{n+1}$.
Thus  the cases with even $n$ are characterized 
by the same reflection amplitude $r=-1$ 
as the free particle, while  for  odd
$n$ we have  $r=+1$.
By this reason if (\ref{HhalfU})
with $U(x)=n(n+1)/x^2$  is considered as 
a Hamiltonian operator in relative coordinate
for a two-particle system,
the cases with even and odd $n$ correspond to bosons
and fermions, respectively.

This picture can be generalized 
to the anyonic case \cite{LeiMyr,Polych}
by considering the same potential
but with coupling constant 
changed for $g=\nu(\nu+1)$ with 
$\nu>0$. The general solution to
the stationary Schr\"odinger equation 
for eigenvalue $E=k^2>0$ with $k>0$ 
is $\psi^{(\nu)}_k(x)=\sqrt{x}\left(C_1J_{\beta}(kx)+C_2Y_{\beta}(kx)\right)$,
where $\beta=\nu+\frac{1}{2}$ 
and $J_\beta$ and $Y_\beta$ are Bessel functions 
of the first and second kind, respectively.
The boundary condition at $x=0^+$ fixes $C_2=0$,
and from asymptotic form of the Bessel function
for $x\rightarrow +\infty$,
$J_\beta(x)=\sqrt{\frac{2}{\pi x}}\cos\left(x-\frac{\beta\pi}{2}
-\frac{\pi}{4}\right)$,
we find that in this case $r=\exp{[-i\pi(\nu+1)]}$,
that generalizes the bosonic and fermionic cases
considered  above for the anyonic case 
of arbitrary $\nu>0$.


\end{document}